\documentclass[sigconf, screen]{acmart}

\usepackage{multirow}
\usepackage{graphicx}
\usepackage{graphics}
\usepackage{adjustbox}
\usepackage{caption}
\usepackage{subcaption}
\usepackage{booktabs,dcolumn}
\usepackage{tabularray}

\usepackage[normalem]{ulem}

\AtBeginDocument{%
  \providecommand\BibTeX{{%
    \normalfont B\kern-0.5em{\scshape i\kern-0.25em b}\kern-0.8em\TeX}}}

\setcopyright{acmcopyright}
\copyrightyear{2024}
\acmYear{2024}
\acmDOI{XXXXXXX.XXXXXXX}

\acmConference[CHI '24]{Make sure to enter the correct
  conference title from your rights confirmation emai}{May 11--16,
  2024}{Honolulu, USA}
\acmBooktitle{CHI '24: ACM CHI Conference on Human Factors in Computing Systems Proceedings, May 11--16, 2024, Honolulu, USA} 
\acmPrice{15.00}
\acmISBN{978-1-4503-XXXX-X/18/06}

\begin{document}

\title[Help Me Reflect: Enhancing Deliberativeness through Diverse Self-Reflection Nudges]{Help Me Reflect: Leveraging Self-Reflection Interface Nudges to Enhance Deliberativeness on Online Deliberation Platforms}

\author{Shun Yi Yeo}
\email{yeoshunyi.sutd@gmail.com}
\affiliation{%
  \institution{Singapore University of Technology and Design}
  \country{Singapore}
}

\author{Gionnieve Lim}
\email{gionnieve_lim@mymail.sutd.edu.sg}
\affiliation{%
  \institution{Singapore University of Technology and Design}
  \country{Singapore}
}

\author{Jie Gao}
\email{gaojie056@gmail.com}
\affiliation{%
  \institution{Singapore University of Technology and Design}
  \country{Singapore}
}

\author{Weiyu Zhang}
\email{weiyu.zhang@nus.edu.sg}
\affiliation{%
  \institution{National University of Singapore}
  \country{Singapore}
}

\author{Simon Tangi Perrault}
\email{perrault.simon@gmail.com}
\affiliation{%
  \institution{Singapore University of Technology and Design}
  \country{Singapore}
}

\renewcommand{\shortauthors}{Shun Yi Yeo, et al.}

\begin{abstract}
The deliberative potential of online platforms has been widely examined. However, little is known about how various interface-based \textbf{reflection nudges impact the quality of deliberation}. This paper presents two user studies with 12 and 120 participants, respectively, to investigate the impacts of different reflective nudges on the quality of deliberation. In the first study, we examined five distinct reflective nudges: \textbf{persona, temporal prompts, analogies and metaphors, cultural prompts and storytelling}. Persona, temporal prompts, and storytelling emerged as the preferred nudges for implementation on online deliberation platforms. In the second study, we assess the impacts of these preferred reflectors more thoroughly. Results revealed a significant positive impact of these reflectors on deliberative quality. Specifically, persona promotes a deliberative environment for balanced and opinionated viewpoints while temporal prompts promote more individualised viewpoints. Our findings suggest that the choice of reflectors can significantly influence the dynamics and shape the nature of online discussions.
\end{abstract}



\begin{teaserfigure}
\centering
  \includegraphics[width=0.78\textwidth]{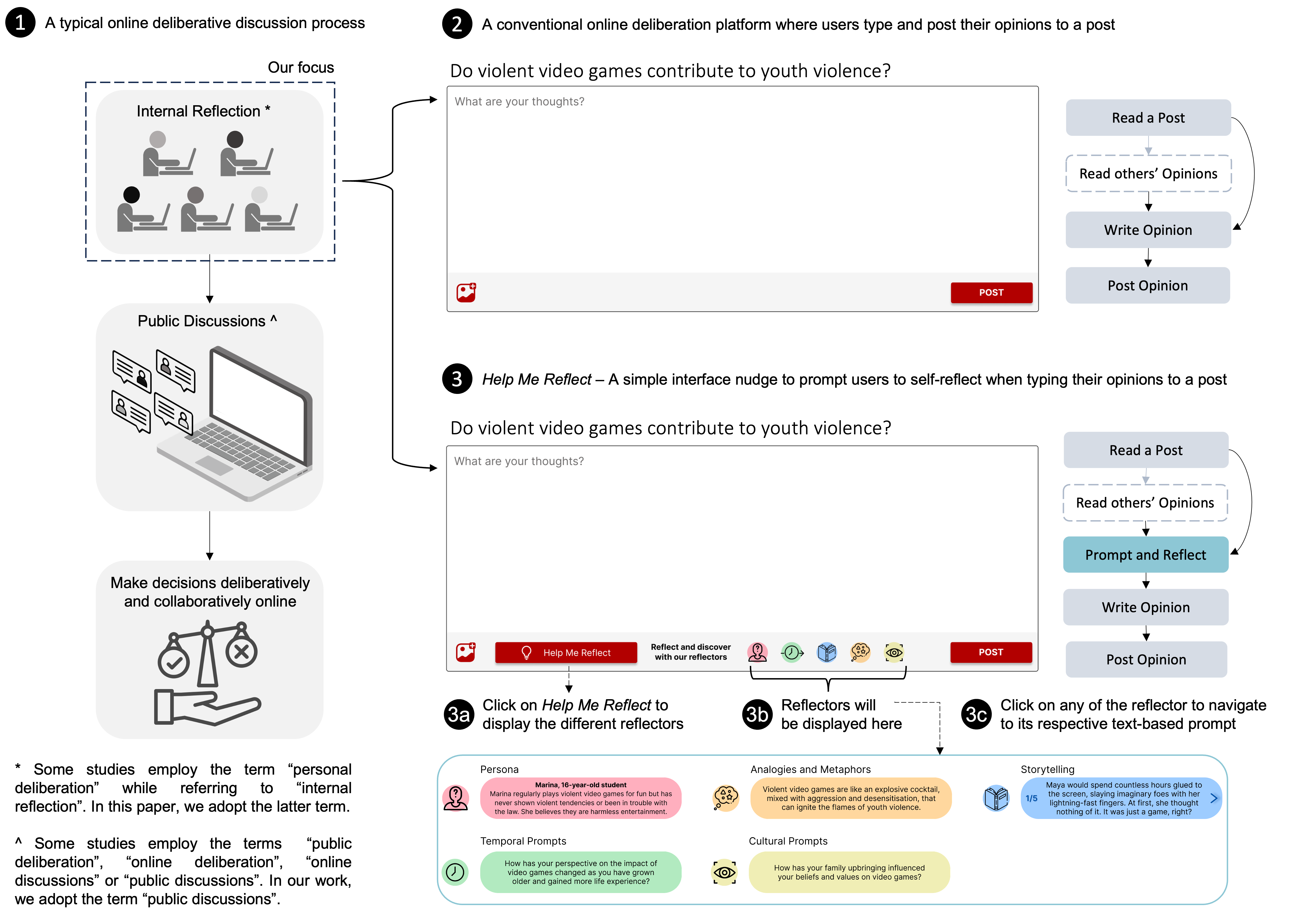}
  \caption{Help Me Reflect: \textbf{1:} A typical online deliberative discussion process involves individuals deliberating independently before collectively discussing and deciding on issues. As such, the quality of ensuing discussions downstream is intricately tied to the quality of internal reflection upstream; \textbf{2:} In a conventional online deliberation platform, users share their opinions on a post with the choice of reading other users' opinions before submitting their own; \textbf{3:} To enhance the quality of public discussion downstream, we introduce \textit{Help Me Reflect}, a simple interface nudge to aid users' self-reflection when they are crafting their comments on a post; \textbf{3a and 3b:} \textit{Help Me Reflect} presents an array of distinct reflectors, each offering a unique approach to guide users' self-reflection through text-based reflective prompts, powered by a Large Language Model: \textbf{3c:} Persona (pink), Temporal prompts (green), Analogies and Metaphors (orange), Cultural prompts (yellow) and Storytelling (blue).}
  \label{fig: teaser}
  \Description{1: Online deliberative discussion process involves individuals deliberating independently before collectively discussing and deciding on issues. Thus, the quality of ensuing discussions downstream is intricately tied to the quality of internal reflection upstream. 2: In a conventional online deliberation platform, users share their opinions on a post with the choice of reading other users' opinions before submitting their own; 3: To enhance the quality of public discussion downstream, we introduce Help Me Reflect, a simple interface nudge to aid users' self-reflection when they are crafting their comments on a post; 3a and 3b: Help Me Reflect presents five distinct reflectors powered by a Large Language Model: 3c: Persona (pink), Temporal Prompts (green), Analogies and Metaphors (orange), Cultural Prompts (yellow) and Storytelling (blue).}
\end{teaserfigure}

\maketitle

\section{Introduction}
Advances in information and communication technologies have enabled the emergence of online forums dedicated to deliberative democracy~\cite{davies2013online, Bertot2010}. Utilizing various formats (e.g., online discourse and reviews, discussion forums and social media), these forums, sometimes called online deliberation platforms~\cite{davies2013online}, enable diverse groups of individuals to contribute their knowledge and opinions~\cite{jacobs2009talking}. Asynchronous exchanges facilitated by discussion threads further empower citizens' participation to collaboratively engage in thoughtful discussions on societal matters~\cite{jacobs2009talking} and informing policy decisions~\cite{dekker2015contingency}. This process is commonly known as online deliberative discussions~\cite{davies2013online, saldivar2019civic} (see Figure~\ref{fig: teaser} (1)). The online deliberation platforms facilitate the process by enabling feasible and well-informed discussions on matters of public significance, with the potential involvement of all stakeholders in exchange of reasons~\cite{zhang2014perceived}.

In some philosophical discussions of deliberation, it is assumed that informed, rational and open-minded discussion will naturally unfold without any further assistance on these online deliberation platforms~\cite{farina2014designing, kim2015factful, habermas1984theory}. However, it is not uncommon to see posts that are inflammatory or off-topic, as well as contributions lacking substantial information~\cite{zhang2005online}. Such common occurrences disrupt the internal reflection process and undermine the quality of subsequent public discussions. Fundamentally, the essence of public discussions is intricately tied to the process of internal reflection, wherein individuals formulate reflective opinions that contribute to the deliberative process~\cite{zhang2021nudge}. This is further reinforced by Goodin and Niemeyer~\cite{goodin2003does} who posit that internal reflection holds a pivotal role in the process of deliberation: \textit{`... the process of one's internal reflection, might be far more important than implied by deliberative democrats’ heavy emphasis on the discursive component.'} Recognizing the primacy of internal reflection, scholarly attention has emphasized reflection as an integral component of deliberation~\cite{bohman2000public, goodin2003reflective}. They argued that fostering reflection is crucial for deliberative democracy to flourish~\cite{muradova2021seeing, chambers2003deliberative, dryzek2002deliberative, goodin2000democratic}. While prior studies have underscored the importance of reflection to augment the quality of deliberation, the comparative analysis of distinct reflection approaches, materialized in interface design, remains largely scant. 

Drawing from the extensive literature on reflection~\cite{yang2021effects}, we seek to provide an examination of various interface-based reflection nudges on the quality of deliberation. Our examination is facilitated by designing and implementing a simple interface design intended to organically induce self-reflection. Our primary focus centers on \textbf{text-based} reflection nudges, to facilitate the integration of the nudges to existing online deliberation platforms. To support our examination in this aspect, we pose the following research questions: \textbf{RQ1: What are the preferred and suitable textual reflective nudges for implementation on online deliberation platforms?} and \textbf{RQ2: How do the different textual reflective nudges affect the quality of deliberation?} We conducted two user studies that each addresses the aforementioned questions respectively.

To answer RQ1, we conducted our first study with 12 participants, where participants interacted with an early prototype of our system, called \textit{Help Me Reflect}, with five selected text-based reflection nudges informed from the literature. The system is powered by a Large Language Model (LLM) (i.e., GPT-3.5\footnote{https://platform.openai.com/docs/models/gpt-3-5}) to provide a range of textual prompts for each of the reflectors.  Based on our interview, we learned which of the five nudges, hereby referred to as reflectors (i.e., persona, temporal prompts, analogies and metaphors, cultural prompts and storytelling), participants preferred and felt more comfortable with. Notably, the top three reflectors were persona, storytelling, and temporal prompts. To answer RQ2, we conducted an experimental study to assess the effectiveness of the different reflectors. We evaluated the system in our experimental study with 120 participants, where we developed three interfaces, each integrating one of the top three reflectors (i.e., persona, temporal prompts and storytelling) to assess their impacts on the quality of deliberation. Study results revealed that all three reflectors trigger self-reflection, leading to higher quality contributions when compared to a control condition. Moreover, the reflectors exhibited distinct traits, leading to varying levels of self-reflection when users formulate their opinions online. We discuss later how the different reflectors influenced the dynamics and environment of online deliberations.   

We make the following contributions: 
\begin{itemize}
    \item The design of integrating self-reflection interface nudges (and using LLM for generalizability and diversification) into online deliberation platforms to enhance deliberativeness.
    \item Identification of the most fitting and preferred self-reflection nudges for implementation on online deliberation platforms from the first study.
    \item An in-depth analysis of the three preferred reflectors, where we highlight the nuanced ways in which each reflective nudge influences the quality and depth of discourse in online deliberation spaces.
\end{itemize}
\section{Background and Related Work} 
We first begin with an overview of our main focus: internal reflection within the deliberation process. Subsequently, we discuss its assessment methods, recent works on nudging to enhance deliberativeness, followed by a review of reflection in conjunction with large language models.

\subsection{The Role of Reflection in Deliberation}
\label{sec: reflection and reflexivity literature review}
Deliberation denotes the process of thoughtful, careful, or lengthy \textit{consideration} by an \textit{individual}~\cite{davies2013online, goodin2003does, aristotle1984complete, hobbes1946}. It fundamentally involves weighing the reasons for and against a given measure~\cite{goodin2003does, oxford}, empowering \textit{individuals} to make informed choices based on due \textit{consideration} of reasons~\cite{christiano2009debates}. Deliberation theorists posit reflection as a fundamental element of the deliberation process~\cite{muradova2021seeing, dryzek2002deliberative, goodin2000democratic, chambers2003deliberative}. Internal reflection inherently precedes public discussion~\cite{goodin2003does} as it is the internal process an individual undergoes that shapes their stance on subsequent public discussions~\cite{goodin2003does}.

The significance of reflection becomes most pronounced when it is used as a standard to evaluate deliberative systems based on their reflective capacity, as proposed by Dryzek~\cite{dryzek2009democratization}. This capacity refers to individuals' ability to \textbf{include others' views and embody views that are of public-mindedness and sincerity}~\cite{holdo2020meta}.
Researchers have explored the use of reflection to enhance deliberation such as through creating a \textbf{pro/con list}~\cite{kriplean2012supporting}, \textbf{perspective-taking} to consider views of various stakeholders’~\cite{kim2019crowdsourcing}, and considering viewpoints of \textbf{an imagined other}~\cite{zhang2021nudge}. Arceneaux and Vander Wielen~\cite{arceneaux2017taming} found that reflection reduces partisan-motivated reasoning and affective polarization. Zhang et al.~\cite{zhang2021nudge} observed how simple \textit{reflective questions} improve attitude clarity and correctness, and encourages opinion expression. Moreover, responding to \textit{reflective prompts} that allow individuals to articulate the rationale behind others' opinions have been demonstrated to enhance deliberativeness~\cite{price2002does}. Collectively, these empirical studies support a shared conclusion: reflection approaches possess the potential to influence deliberativeness, a normatively desired effect for deliberation~\cite{bohman2000public, zhang2021nudge}. 

\begin{figure*}[!htbp]
  \centering
  \includegraphics[width=\linewidth]{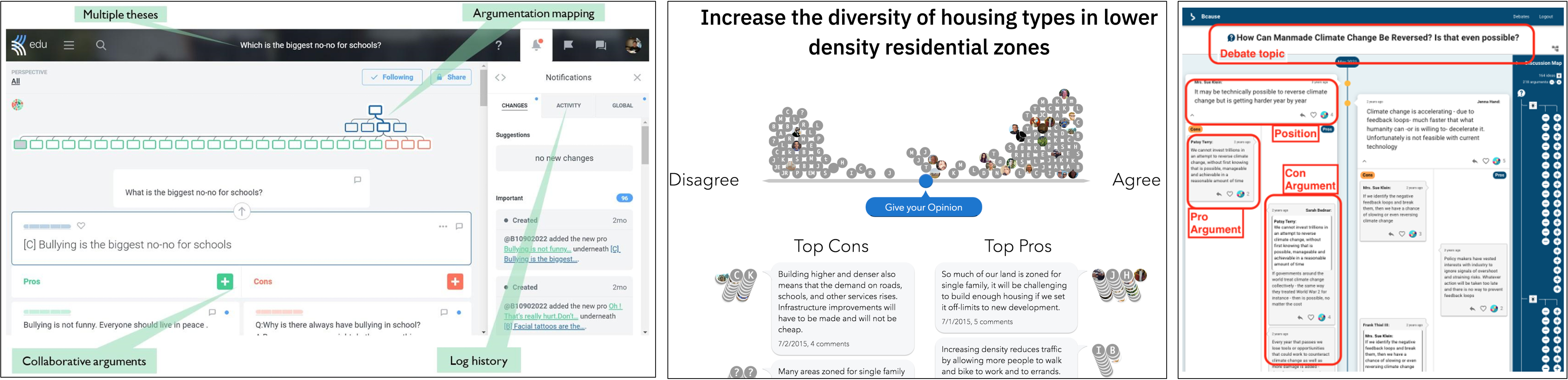}
  \caption{Examples of existing online deliberation platforms that integrate reflection approaches (from left): Darmawansah et al.~\cite{darmawansah2022empowering} implemented the \textbf{Collective Reflection-based Argumentation Mapping (CR-AM)} strategy on online platform, Kialo Edu. On this platform, students collaboratively respond to a specific thesis by providing pro/con claims, supporting arguments and counterarguments; \textbf{ConsiderIt}~\cite{kriplean2012supporting} is a platform to support users' reflection process by guiding them to reflect on trade-offs of policies through the creation of pros and cons points; \textbf{BCause}~\cite{anastasiou2023bcause} incorporates reflective approaches by using a pro/con argumentation structure to organize the conversation and facilitate healthy online deliberation.}
  \label{fig: existing online deliberation platforms}
  \Description{A figure depicting the interfaces of three examples of existing online deliberation platforms that incorporates reflection approaches. From left, Darmawansah et al. implemented the Collective Reflection-based Argumentation Mapping (CR-AM) strategy on online platform, Kialo Edu, where students jointly respond to a specific thesis by providing pro/con claims, supporting arguments and counterarguments. Next, ConsiderIt is a platform to support users' reflection process by guiding them to reflect on trade-offs of policies through the creation of pros and cons points. Lastly, BCause is a platform that incorporates reflective approaches by using a pro/con argumentation structure to organize the conversation and facilitate healthy online deliberation.} 
\end{figure*}

Investigations to enhance deliberation quality through reflection have been widely examined as depicted above. However, the influence of different reflection approaches on deliberativeness remains inadequately comprehended. Without this knowledge, uninformed use of reflection approaches may result in superficial utilization and conflicting outcomes~\cite{vink2022building, akama2013embodying}. Furthermore, not scrutinizing their varying effects creates the risk of favoring one reflection practice over others. As the uptake of integrating reflection approaches grows on deliberation platforms~\cite{ercan2019public, goodin2003does, anastasiou2023bcause, darmawansah2022empowering, kim2019crowdsourcing, klein2011harvest} (see Figure~\ref{fig: existing online deliberation platforms}), it becomes imperative to support this facet.

For this study, we thus focus on examining and comparing the efficacy of a variety of reflection approaches in the context of online deliberation platforms. We seek to contribute to the understanding of distinct approaches of reflection and their effects on the deliberative process. Consequently, we expect this study to provide a more thoughtful and comparative use of reflective nudges to support deliberation. 

\subsection{Measuring Deliberation: Deliberativeness}
\label{sec: measurements}
Deliberativeness denotes the quality of deliberation made by an individual~\cite{trenel2004measuring}, a core value of public discussions~\cite{menon2020nudge}. In our study, we adopt the term `deliberativeness' to characterize the quality of an individual's opinion~\cite{price2002does, bohman2000public, zhang2021nudge, steenbergen2003measuring, stromer2007measuring, graham2003search} on online deliberation platforms.

Deliberativeness is a composite measure with various dimensions~\cite{trenel2004measuring, zhang2021nudge}. Trénel~\cite{trenel2004measuring} proposed a coding scheme for deliberativeness with eight measurements, of which two dimensions: \textit{rationality} and \textit{constructiveness} constitute the core conditions of deliberation. Rationality comprises two components: \textit{opinion expression}, which examines whether opinions are articulated~\cite{trenel2004measuring}, as deliberative processes necessitate the presence of opinion statements~\cite{trenel2004measuring}; and \textit{justification level}, which captures the degree of reasoning to justify and support one's claim~\cite{trenel2004measuring}. Constructiveness captures the degree of advancing considerations beyond oneself~\cite{kymlicka2002contemporary}, emphasizing the orientation towards finding common ground~\cite{trenel2004measuring, goddard2023textual} . Trénel measures the constructive nature of an opinion by its orientation of mentioning others' interests, specifically whether the opinion is one-sided~\cite{trenel2004measuring} after an inclusive consideration of diverse viewpoints~\cite{black2011self, gastil2007public}. In the philosophies of Plato and Aristotle~\cite{walton1999one}, a constructive, two-sided argument necessitates a balanced attitude. Niculae and Niculescu-Mizil~\cite{niculae2016conversational} also showed that contributions characterized by higher constructiveness tend to be more well-balanced.
Therefore, both rationality and constructiveness are central to the process of deliberation.

In addition, Cappella et al.~\cite{cappella2002argument} propose \textit{argument repertoire} which quantifies the quality of participants' opinions by counting the number of non-redundant arguments for and against an issue, a common measure used in deliberative research~\cite{menon2020nudge, zhang2021nudge, cappella2002argument, kim2021starrythoughts}. Other papers also emphasize the significance of \textit{argument diversity}~\cite{anderson2016all, gao2023coaicoder, richards2018practical}, which measures the diversity of perspectives in an opinion, accounting for varied interpretations and viewpoints of an issue. 

For this study, we follow the practice of prior work, assessing deliberativeness with five measures: rationality (opinion expression and justification level), constructiveness, argument repertoire and argument diversity. 

\begin{figure*}[!htbp]
  \centering
  \includegraphics[width=\textwidth]{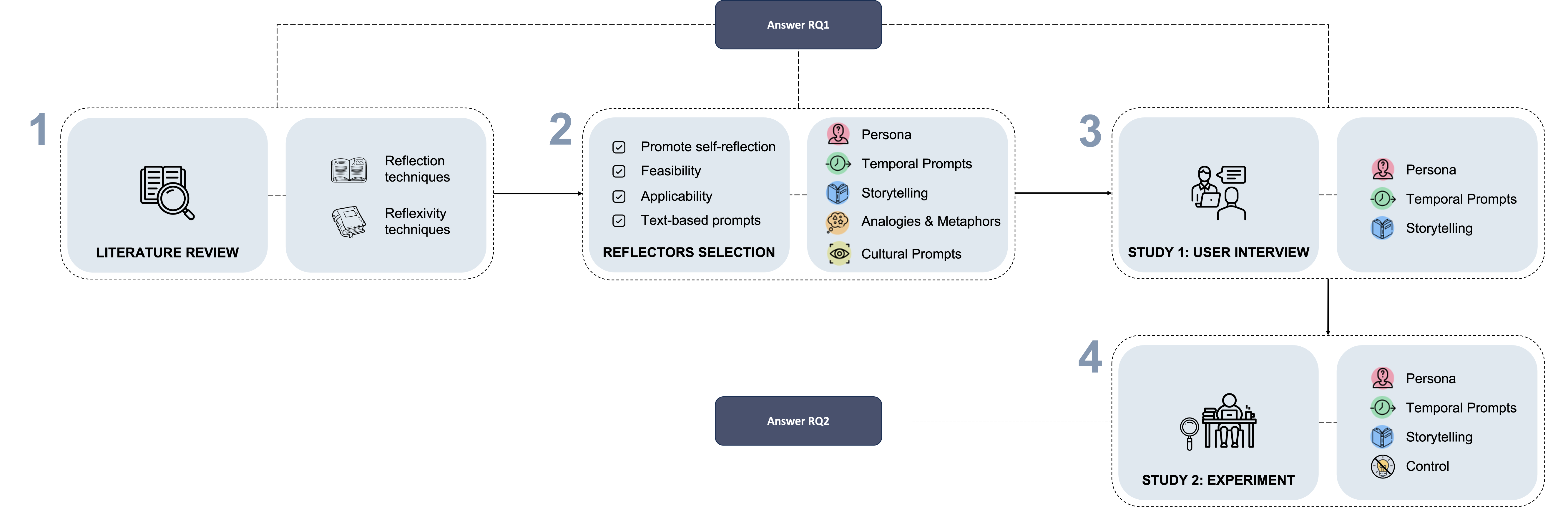}
  \caption{Study Procedure. \textbf{1:} We reviewed existing literature on reflection and reflexivity approaches on self-reflection for online deliberation. \textbf{2:} We excluded approaches that do not align with our selection criteria. \textbf{3.} We then conducted our first user study to answer RQ1 on users' preferred reflectors for implementation on online deliberation platforms. \textbf{4.} Subsequently, we conducted an experimental study to answer RQ2 on assessing the impacts of the reflectors on users' deliberativeness.}
  \Description{Study Procedure. 1: We reviewed existing literature on reflection and reflexivity approaches on self-reflection for online deliberation. 2: We excluded approaches that do not align with our selection criteria: (a) promote self-reflection, (b) feasibility, (c) applicability and (d) text-based prompts. 3. We then conducted our first user study to answer RQ1 on users' preferred reflectors for implementation on online deliberation platforms. The first three procedures answer RQ1. 4: Subsequently, we conducted an experimental study to answer RQ2 on assessing the impacts of the reflectors on users' deliberativeness.} 
  \label{fig: study procedure}
\end{figure*}

\subsection{Nudging to Enhance Deliberativeness}
\label{sec: nudging}
Nudges involving simple interface changes have been shown to effectively change users' behaviour~\cite{thaler2008nudge}. Nudges seek to influence choices and decision process by effectively intervening and moderating an individual's usual behavior~\cite{thaler2008nudge}, redirecting them to shift from their habitual mode of thought towards the value introduced by the nudge~\cite{thaler2008nudge}. 

For instance, Wang et al.~\cite{wang2014field} found that the sense of `regret' experienced during online disclosures on Facebook could be lessened by using behavioral strategies aimed at heightening users' privacy awareness such as visual cues and timer interventions.

Within the realm of deliberation, there has been some interest in using simple interface designs to enhance the deliberativeness of online discourse~\cite{xiao2015design, zhang2013structural}. For instance, Menon et al.~\cite{menon2020nudge} proposed three interface nudges: word-count anchor, partitioning text fields and reply choice prompt. The authors showed that small interface tweaking like the partitioned text fields can lead to longer replies by up to 35\% and more arguments by up to 25\%. Also, Zhang et al.~\cite{zhang2021nudge} examined the use of reflection nudges through simple question prompts on users' perceived issue knowledge, attitudes and opinions expression. Their findings showed that asking participants to answer questions like `what are your opinions on this issue?' enhanced the quality of participants' opinions. Our work differs from Zhang et al.~\cite{zhang2021nudge} by examining a wider variety of reflective nudges and their impacts on different dimensions of deliberative quality.

Drawing from existing efforts, we utilize the concept of nudges to trigger reflection organically. Our goal is to enhance the deliberative quality through various interface-based reflection nudges (see section~\ref{sec: reflection and reflexivity literature review}). We anticipate that these reflective nudges will enhance deliberativeness. 

\subsection{Reflection and Large Language Models}
\label{sec: reflection and LLM}

Large language models (LLMs) have become integral to human communication~\cite{bommasani2021opportunities}. These models are capable of generating human-like language~\cite{jakesch2023human} and have been applied in various domains like writing assistance~\cite{dang2022beyond, jakesch2023co, hancock2020ai}, and grammar support~\cite{koltovskaia2020student}. As we incorporate these models into our daily interactions, they have the potential to not only shape our behaviors but also to influence our opinions~\cite{jakesch2023co}. In one study, the authors called this new paradigm of influence, \textit{latent persuasion}~\cite{jakesch2023co} - a term that draws parallels with the concept of nudge (see section~\ref{sec: nudging}), but now extended into the realm of machine-generated language and persuasion~\cite{fogg2002persuasive, leonard2008richard}. 

Research on LLM and Natural Language Processing (NLP) have shed light on their role as active partners throughout the writing process~\cite{lee2022coauthor, yuan2022wordcraft}. These studies have extensively explored various aspects of interface design and user interactions with writing assistants to improve language proficiency ~\cite{buschek2021impact}, facilitate story creation~\cite{singh2022hide, yuan2022wordcraft}, aid in text revision~\cite{cui2020justcorrect, zhang2019type}, and foster creative writing~\cite{clark2018creative, gero2019metaphoria}. In many cases, these investigations have positioned language models as coauthors, active collaborators, or active writing assistants~\cite{lee2022coauthor, yang2022ai}, all aimed at assisting users in improving their writing skills. However, as larger and more powerful language models continue to emerge~\cite{bommasani2021opportunities}, the majority of research has centered around the potential utilization of these models to influence and assist users in their writing. There remains a notable gap where these language models can be harnessed to stimulate users' reflection, a crucial consideration since our reflective thinking often shapes our writing~\cite{kellogg1999psychology}. For this work, we thus leverage on LLM (i.e., GPT-3.5) as a support to stimulate users' reflection, helping them in their deliberation process when writing their opinions on an issue. 
\section{Selecting the Reflectors}
\label{sec: reflectors selection}

\begin{figure*}[!ht]
  \centering
  \includegraphics[width=\linewidth]{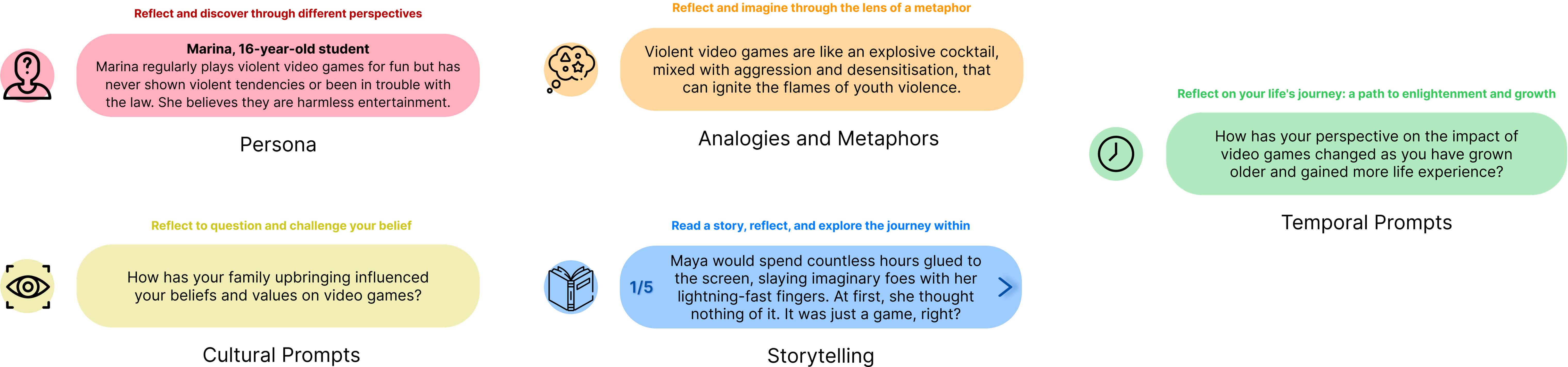}
  \caption{Example of the Five Selected Reflectors.}
  \label{fig: reflector description}
  \Description{Example of the five selected reflectors. An example of persona: Reflect and discover through different perspectives - Marina, 16-year-old student. Marina regularly plays violent video games for fun but has never shown violent tendencies or been in trouble with the law. She believes they are harmless entertainment. An example of analogy and metaphor: Reflect and imagine through the lens of a metaphor - Violent video games are like an explosive cocktail, mixed with aggression and desensitisation, that can ignite the flames of youth violence. An example of cultural prompt: Reflect to question and challenge your belief - How  has your family upbringing influenced your beliefs and values on video games? An example of storytelling: Read a story, reflect, and explore the journey within - Story page 1 out of 5. Maya would spend countless hours glued to the screen, slaying imaginary foes with her lighting-fast fingers. At first, she thought nothing of it. It was just a game, right? An example of temporal prompt: Reflect on your life's journey: a path to enlightenment and growth - How has your perspective on the impact of video games changed as you have grown older and gained more life experience?} 
\end{figure*}

Our goal is to organically nudge for self-reflection through different reflection approaches (i.e., reflectors). As such, we first rely on literature on reflection and reflexivity as important theoretical inputs when identifying the designated reflectors (see Figure~\ref{fig: study procedure} (1 and 2)). A comprehensive exploration of pertinent literature was conducted as detailed in section~\ref{sec: reflection and reflexivity literature review}. Notably, our selection of reflectors were largely informed by Vink and Koskela-Huotari~\cite{vink2022building}, and papers from HCI domains, particularly those directed towards augmenting deliberative discourse on online platforms. Other literature were from the broader realm of social sciences to ensure a holistic comprehension of the identified reflectors. 

We employed a snowball search method~\cite{wohlin2014guidelines} to facilitate the targeted acquisition of papers that explicitly centered on reflection within the domain of online deliberation. Leveraging references cited within these papers, we subsequently identified other relevant studies. To qualify for inclusion within our literature study, each paper had to overtly address various types of reflection and/or reflexivity or concentrate on the enhancement of deliberation. We made these determinations by reviewing abstracts or reading full articles as needed.

When selecting the reflection approaches, we adhered to four key criteria: 
\begin{itemize}
    \item (C1) their capacity to promote self-reflection among users; 
    \item (C2) feasibility for implementation within online deliberation platforms; 
    \item (C3) applicability across a broad spectrum of online discussion topics;
    \item (C4) text-based prompts to align with the text-focused nature of current online deliberation platforms.
\end{itemize}
For instance, we excluded reflection approaches that center on bodily and sensory experiences as catalysts for self-reflection~\cite{creed2020institutional, rigg2018somatic} (C2). We also omitted reflection approaches that center on visuals (e.g., pictures and videos) as users primarily express their opinions in the form of textual comments on online forums like Reddit and on most online deliberation platforms (C4).

As a result, we identified five distinct reflection nudges suitable for integration into online deliberation platforms: 
\begin{enumerate}
    \item \emph{Persona} follows the design principles as Cooper~\cite{cooper1999inmates, cooper2007face, pruitt2010persona}, who initially introduced the concept within the HCI community. It functions as a fictional character embodying distinct demographics (i.e., age and occupation), aiming to stimulate users' self-reflection from varied viewpoints. It has found applications in various works~\cite{kim2019crowdsourcing, zhang2021nudge, vink2022building}, emphasizing the perspective-taking approach as a means to encourage self-reflection. 
    \item \emph{Analogies and metaphors} follows the design principles outlined by scholarly definitions~\cite{Webster,gentner1982scientific,lee2007causal}. They serve as figurative speech tools designed to assist individuals in re-framing their thoughts. This approach has been effectively utilized in various studies to support cognitive reflection~\cite{vink2022building, keefer2016metaphor, schwind2009metaphor, niebert2012understanding}.
    \item \emph{Cultural Prompts} seek to cultivate self-awareness through reflective questions that encourage individuals to reflect on their beliefs, biases and personal context~\cite{adams2003reflexive, donati2011modernization, mouzelis2010self}. This approach has found most applications in fostering self-reflection among learners and educators~\cite{civitillo2019interplay}. 
    \item \emph{Storytelling} seek to stimulate self-reflection through short narratives, an approach that has been observed to promote introspection and improve individuals' ability to articulate their thoughts~\cite{freidus2002digital}.  
    \item \emph{Temporal Prompts} encourage self-reflection by prompting individuals to reflect on their personal history, life experiences and past events~\cite{phemister2017leibniz, vink2022building}, an approach that was employed in various studies~\cite{lindstrom2006affective, smallwood2011self, murakami2017connecting, phemister2017leibniz}. 
\end{enumerate}

Figure~\ref{fig: reflector description} provides examples for each of the different selected reflectors.

The above five nudges correspond to the promotion of self-reflection through relational, cognitive, temporal and cultural approaches, respectively, as outlined by Vink and Koskela-Huotari~\cite{vink2022building}. For the full description of the selected reflectors, refer to Appendix Table~\ref{tab: reflectors selection}. The selection of these five nudges is aimed at maintaining a representative set that encapsulates the diversity of reflective approaches from the literature. 

\subsection{Implementation of the Reflectors with LLMs}
For implementation, the five nudges were implemented by leveraging the capabilities of LLMs (i.e., OpenAI's ChatGPT model, GPT-3.5) to generate a range of textual prompts for each reflector. Specifically, we instructed the LLM to generate six distinct textual prompts for each reflector, resulting in a total of 30 textual prompts across the five reflectors. The utilization of LLMs to promote self-reflection on online deliberation platforms ensures the adaptability and scalability of the reflectors. Using LLMs, we tap into its vast knowledge and language proficiency in enabling the reflectors to accommodate to a diverse array of topics commonly found in online discussions. This enhances the generalizability and versatility of the reflectors to effectively cater to the varied and dynamic nature of discussions on online deliberation platforms.

In creating the textual prompts, GPT is tasked with the role of ``\textit{a helpful assistant focusing on supporting users' self-reflection on a given topic}''. We also tailored different prompts for the distinct reflectors. Specifically, we followed the design principles established by scholarly definitions for each reflector, ensuring alignment with the intended objective of each reflective nudge. Additionally, we followed the prompt template by White et al.~\cite{white2023prompt} when prompting GPT: define a task, add constraints and set out expectations. For example, for the persona reflector, we prompt the model with ``\textit{Create six distinct personas representing different perspectives on the topic. Provide the name, age and occupation for each persona.}'' Specific constraints were also established for the persona reflector: ``\textit{Create three male personas and three female personas.}'' This constraint was set out as scholars have found that LLMs such as GPT-3 produce gender stereotypes and biases~\cite{brown2020language,lucy2021gender, huang2019reducing, nozza2021honest, johnson2022ghost}. Hence, this constraint seek to mitigate any potential gender imbalances for the perspectives that were generated by GPT. All prompts, along with the rationales behind the constraints set out for GPT, are listed in Appendix Table~\ref{tab: prompt engineering} and ~\ref{tab: prompt constraint rationale}. Appendix Figure~\ref{fig: GPT persona output} shows an example output of the textual prompts generated by GPT. To ensure diversity of the textual prompts without being overly random, the temperature parameter is set at 0.7.
\section{Study 1: Determining the Preferred Reflectors}
\label{sec: study 1}
To address \textbf{RQ1} and discern the preferred and suitable reflectors for integration into online deliberation platforms, we conducted a user study followed by semi-structured interviews with the participants ($N=12$), focusing on their inclinations towards specific reflectors and the manner in which each reflector contributed to their self-reflection within an online deliberation context. Details of the five reflectors are outlined in section \ref{sec: reflectors selection}. 

\subsection{Independent and Dependent Variables}
A within-subject experiment was conducted with one independent variable being the Reflector with five levels: \{Persona, Analogy and Metaphor, Cultural Prompts, Storytelling, Temporal prompts\}. The order of presentation of the reflectors was randomized for each participant to counterbalance any ordering effect.
The study has one dependent variable, Ranking, to discern users' preferences for the reflectors.

\subsection{Apparatus}
The study was conducted remotely using Zoom. Participants were instructed to access a simulated online deliberation environment and to share their screen.
This environment was based on Reddit, a prominent global online discussion platform~\cite{horne2017identifying, medvedev2019anatomy}, to evaluate the reflectors within a context that closely aligns with realistic and relevant settings.

\subsubsection{Implementation}
The prototype was developed using Figma\footnote{https://www.figma.com/} and ProtoPie\footnote{https://www.protopie.io/} before its deployment on Useberry\footnote{https://www.useberry.com/}, a user testing site.
We developed the feature - \textit{Help Me Reflect} - which displays the five reflectors selected for the study.

\subsubsection{Key Features of the Prototype}
\begin{figure*}[!htbp]
  \centering
  \includegraphics[width=\textwidth]{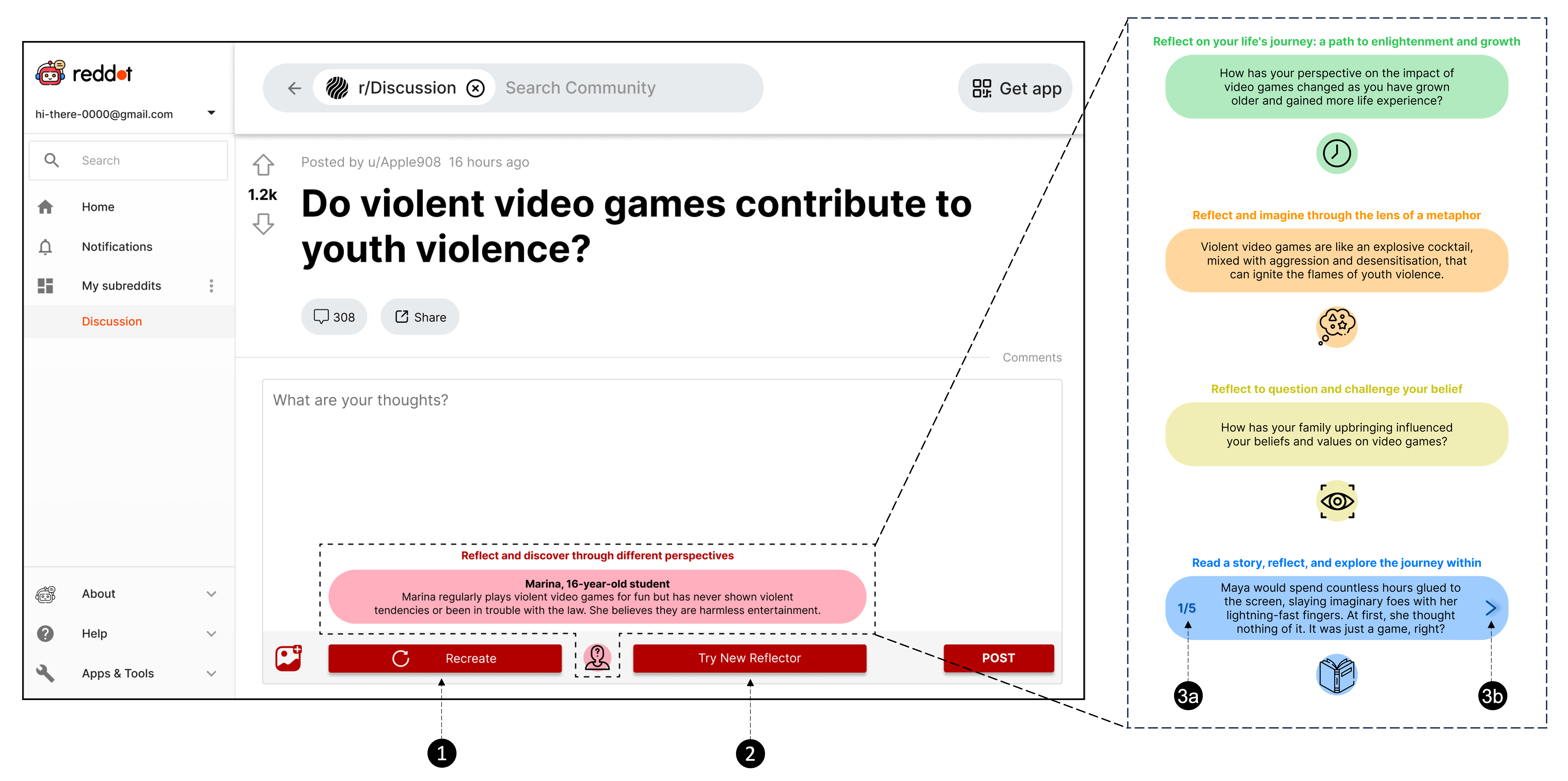}
  \caption{Key features in Help Me Reflect. \textbf{1:} Users can click on the \textit{Recreate} button to browse through different textual prompts under a specific reflector or \textbf{2:} click on the \textit{Try New Reflector} button to navigate back to the different array of reflectors to choose another reflector. Additional features are present for storytelling to allow users to \textbf{3a:} track their reading process and \textbf{3b:} continue with their reading on the current story.}
  \label{fig: features}
  \Description{The user interface created for the study encompasses the following features in the comment box: 1: Users can click on the Recreate button to browse through different textual prompts under a specific reflector or 2: click on the Try New Reflector button to navigate back to the different array of reflectors to choose another reflector. When clicking on the storytelling reflector, additional features are present: 3a: The numbers, for example, page 1 out of 5 allows users to track their reading process. 3b: The forward arrow allows users to click and continue with their reading on the current story.}
\end{figure*}

To facilitate users' self-reflection during the writing process on a discussion topic, we designed several key features. These are depicted in Figures~\ref{fig: teaser} and \ref{fig: features}.

\paragraph{Help Me Reflect.} \textit{Help Me Reflect} is a feature that can be integrated into online discussion platforms to facilitate users in their self-reflection process when they are crafting their opinions on a discussion topic. Clicking on \textit{Help Me Reflect} displays the designated reflectors. Users can choose a specific reflector, allowing them to explore and delve deeper within the context of that chosen reflector. These are depicted in Figure~\ref{fig: teaser} (3a and 3b). 

\paragraph{Recreate} Within that chosen reflector, users can click on \textit{Recreate} to browse through different textual prompts available for the current reflector. See Figure~\ref{fig: features} (1). In the case of persona and storytelling, we also ensured that the textual prompts alternate between males and females, as well as between positive and negative tones. This approach was implemented to maintain a balanced sequence, preventing any undue one-sidedness.

\paragraph{Try New Reflector.} To choose a different reflector, users can click on \textit{Try New Reflector} to navigate back to the array of reflectors. See Figure~\ref{fig: features} (2).

\paragraph{Features Specific to Storytelling.}
While reading a story, users can track their progress using the page tracker displayed in Figure~\ref{fig: features} (3a), which indicates the number of remaining pages until the story concludes. Short stories comprises of four to five pages, medium length stories have six to seven pages and long stories have eight to nine pages. To continue reading the ongoing story, users can simply click on the \textbf{\huge{>}} button (Figure~\ref{fig: features} (3b)). 

\subsection{Participants and Ethics}
After obtaining approval from our local Institutional Review Board (IRB), we recruited 12 participants (four males and eight females), with an average age of 23.3 ($SD=4.86$), following the local guidelines for sample size~\cite{caine2016local}. Notably, interviews conducted in remote settings have a mean sample size of 15 participants with $SD=6$, while the most common sample size is 12 in manuscripts published at CHI~\cite{caine2016local}.
The participants were university students with the complete demographic information available in Appendix Table~\ref{tab: st1-demo}.

On average, participants took \textbf{66.9 minutes} to complete the entire study. Compensation for participants was set at a rate equivalent to 7.25 USD per hour in accordance with our IRB.

\subsection{Task and Material}
\label{sec:maintask}
For the task, we identified a discussion topic through ProCon.org\footnote{https://www.procon.org/}, an online resource for research on contentious issues. The selected topic was “Do violent video games contribute to youth violence?”, a topic that is readily accessible and politically relevant, thereby promoting constructive and open debate.

In the main task, participants used our prototype with the \textit{Help Me Reflect} feature. \textbf{Participants had to write at least 30 words with the help of each of the five reflectors when expressing their take on the discussion topic.} During this process, participants engaged with all five reflectors sequentially (as the study was conducted over Zoom, the researcher instructed and guaranteed this implementation). The names of the reflectors were deliberately withheld with only the icons shown (as depicted in Figure~\ref{fig: teaser} (3b)). This was done to prevent participants from adapting their behavior based on their knowledge of the reflectors' functions and to ensure that they engaged with the reflectors authentically and without preconceived notions of their intended effects.

There were also no hard rules dictating the manner in which participants should utilize \textit{Help Me Reflect}. They could concurrently write while using the reflectors for real-time self-reflection, initially employ the reflectors for reflection before composing their opinions, or first formulate their opinions and then integrate reflections facilitated by the reflectors. Furthermore, \textbf{participants were not obliged to explore every textual prompt within a particular reflector, though they had the option to do so if desired.}

\subsection{Procedure}
\label{procedure}
The study contained three parts with Useberry and Qualtrics being used for data collection.

\subsubsection{Pre-Task: Consent, Instructions, Personal Information, Topic Knowledge and Interest, Self-Reflection Questionnaire}
Participation consent was obtained before the study. Participants were informed that they had to express their opinions on a discussion topic, which was undisclosed at this point, using various reflectors. They then provided general demographic information.

Given the discussion topic on video games, we considered dimensions like \textbf{topic knowledge and topic interest (TK-TI)} to assess participant variability on the topic. These dimensions were evaluated through a six-item multiple choice/fill-in-the-blank questionnaire presented in the Appendix Figure~\ref{fig: TK-TI Questionnaire}. Each item was scored 1 if participants displayed knowledge or interest and 0 otherwise, thus giving a final score between 0 and 6. Our participants scored an average of ($M=3.58$, $SD=2.71$).

We also administered a \textbf{Self-Reflection and Insight Scale (SRIS) questionnaire}~\cite{grant2002self}. The SRIS is a widely used self-reported scale that assesses individual differences in their capacity and tendency to self-reflect~\cite{silvia2022self}. Grounded in models of meta-cognition and personal development~\cite{grant2001rethinking,grant2003impact}, the SRIS employs a 20 questions on a 1-6 Likert scale (1 = strongly disagree to 6 = strongly agree) to evaluate self-reflection and insight. The SRIS maximum score is 120, with participants' scores ranging from 75 to 119 ($M=91.5$, $SD=13.7$). We included these controlled variables in the analysis by using ANCOVA instead of ANOVA. 

\subsubsection{Main Task} This has been detailed in section~\ref{sec:maintask}.

\subsubsection{Post-Task Interview}
After completing the main task, we conducted semi-structured interviews to gather feedback, and also asked participants to rank the five reflectors from 1 (lowest) to 5 (highest) based on their personal preference and the extent of self-reflection induced by each of the reflectors. Participants were asked to explain the reasons behind their ranking. We also asked questions specifically relating to the challenges they faced for each reflector and the helpfulness of the \textit{Help Me Reflect} feature in their thought process. All interviews, lasting between 15 to 35 minutes, were audio recorded and transcribed through Zoom to facilitate subsequent analysis. 

\subsection{Findings}

\subsubsection{Ranking}

\begin{figure*}[!htbp]
  \centering
  \includegraphics[width=.6\textwidth]{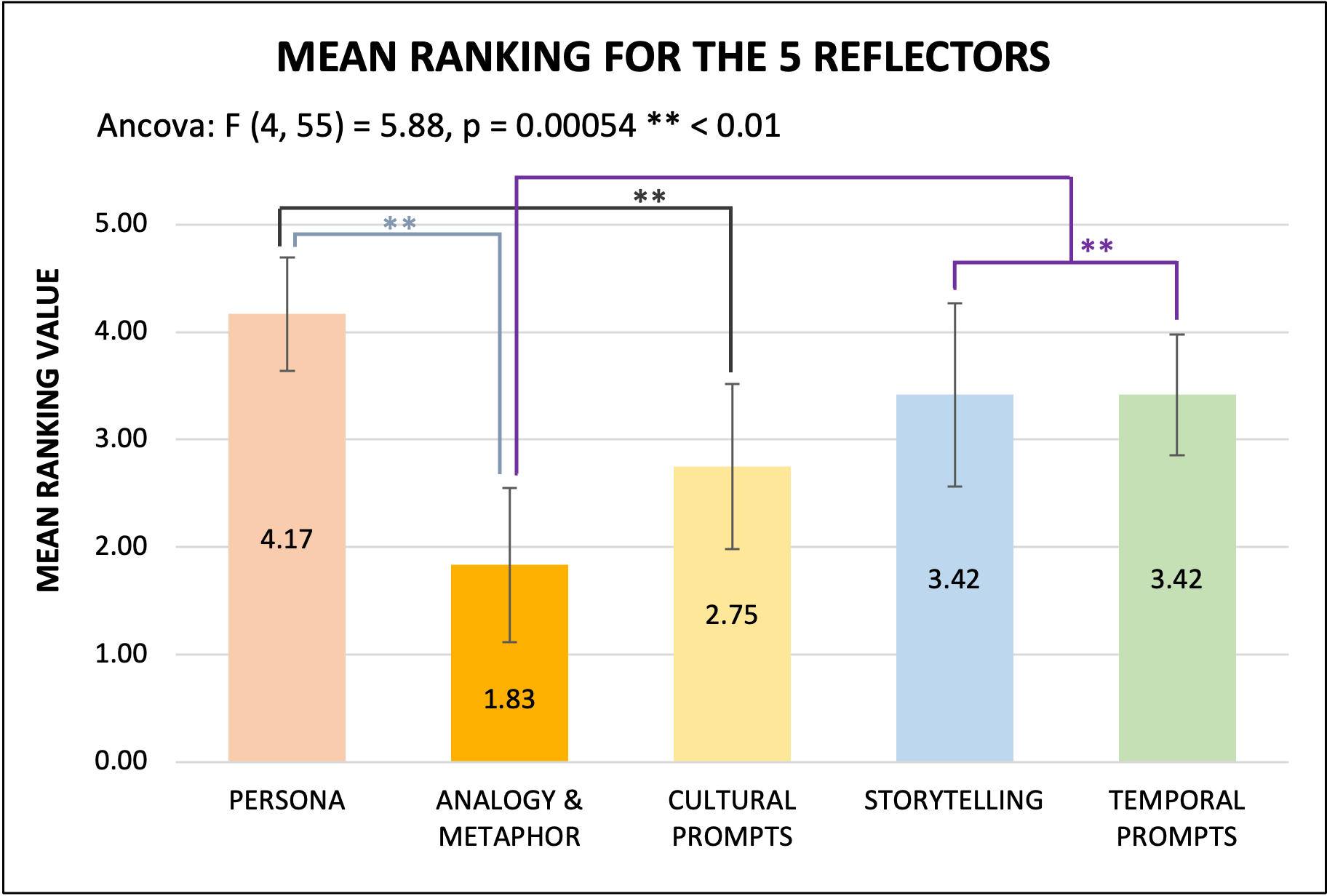}
  \caption{Mean ranking for the five reflectors with 1 being the lowest rank and 5 being the highest rank. Error bars show .95 confidence intervals. We report the results of the ANCOVA test and pairwise comparisons with BH correction, where * : p < .05, ** : p < .01.}
  \label{fig: ranking}
  \Description{Bar graph describing the mean rankings for the five reflectors. The X-axis shows the different reflectors and the Y-axis shows the ranking value ranging from 0 to 5 at a rank interval of 1. 1 is the lowest possible rank and 5 is the highest possible rank. The highest mean ranking was found in persona at 4.17, followed by storytelling and temporal prompts at 3.42. Cultural prompts received a mean ranking of 2.75, while analogy and metaphor had the lowest ranking at 1.83.}
\end{figure*}

Figure~\ref{fig: ranking} shows the average rankings for each reflector. On average, the \emph{Persona} reflector received the highest ranking with a score of 4.17 out of 5, while \emph{Analogies and Metaphors} reflector received the lowest ranking with an average score of 1.83 out of 5.
A one-way repeated measures ANCOVA was conducted to determine a statistically significant difference between the Reflectors on Ranking, controlling for covariates TK-TI and SRIS (see section~\ref{procedure} for the controlled variables). We found a statistically significant main effect of the Reflectors on Ranking ($F_{4,55} = 5.88$, $p<.01$). Specifically, the \emph{Persona} reflector was significantly ranked higher compared to \emph{Analogies and Metaphors} ($p<.01$) and \emph{Cultural Prompts} ($p<.01$). \emph{Storytelling} and \emph{Temporal Prompts} were also ranked significantly higher than \emph{Analogies and Metaphors} (both $p<.01$). We did not find any effect of both TK-TI and SRIS scores on Ranking. In conclusion, the persona, storytelling, and temporal prompts were the top three reflectors.

\subsubsection{Qualitative Feedback}
We provide a comprehensive account of participants' experiences with each of the reflectors, highlighting both positive aspects and challenges encountered. The notation (/12) indicates the count of participants who shared the same observation. Additionally, we also detail further insights participants offered on the \textit{Help Me Reflect} feature.

\paragraph{Persona}
Participants appreciated the persona reflector as it encouraged them to consider alternative viewpoints and diverse perspectives (7/12). ``\textit{I felt that the persona made me think of the other side. Whereas when I first got the answer I had in mind for the discussion topic, I just didn't look at the other side or think of other perspectives. I thought looking at the other side is very important, that's why I rank that the first.}'' (P3). 

The persona reflector was also praised for facilitating a deeper understanding of others' viewpoints: ``\textit{This reflector allows me to see through different people's perspectives. It is more helpful in knowing how other people think than just me asking myself `what do I think about the issue'.}'' (P10). Additionally, participants appreciated the specificity and clarity provided by this reflector: ``\textit{This reflector is good because it provides a name, age, as well as the experience of the person. I like how it is specific and easy to understand so that I can quickly understand how others perceive the topic.}'' (P11). Finally, it
also engendered a sense of familiarity for participants, akin to real-life conversations (2/12): ``\textit{This reflector makes me feel like I am talking to a friend who is sharing his/her own belief on a topic. In that sense, I am more open to understanding others' perspectives.}'' (P4).

\paragraph{Analogy and Metaphor}
\label{sec: analogy and metaphor findings}
Compared to other reflectors which offer simple information or questions, the analogies and metaphors system adds a layer of abstractness making it seemingly harder for participants to reflect. Similarly, as metaphors cannot be 100\% accurate in terms of representation, participants seemed to focus on the less accurate aspects.

As such, it received mixed feedback. Only one participant found this reflector interesting and relatable: ``\textit{I thought the analogies and metaphors are quite interesting. When we use metaphors, they're usually adapted from real-life examples. So to me, it's something I can relate to.}'' (P2). Two participants viewed this reflector humorously (2/12): ``\textit{It's just very cringe - I started laughing when I read the metaphors. It's more like a joke.}'' (P4) Another two expressed dissatisfaction with its effectiveness (2/12): ``\textit{The metaphors don't make any sense to me.}'' (P1). There were concerns about this reflector seeming to be sometimes one-sided (2/12): ``\textit{I don't quite agree with the metaphors because all issues have two sides - there are different factors that may or may not trigger violence.}'' (P6). In addition, participants felt that they require prior knowledge to understand the reflector which hindered self-reflection (3/12): ``\textit{The reader needs to have background or cultural knowledge to understand. For me, I don't have this cultural background to comprehend the metaphors.}'' (P11). 

\paragraph{Cultural Prompts}
The cultural prompts reflector elicited varied responses from the participants. Some found it challenging to relate it to themselves (4/12): ``\textit{Certain things in the questions don't really relate to me - for example, I don't see how beliefs and family upbringing can relate to video games.}'' (P4). Some participants noted the cognitive burden imposed by this reflector (2/12): ``\textit{This reflector has a lot of rhetorical questions, so I need to spend more time thinking. When I see things online, I don't want to go through the whole process of thinking.}'' (P10). Another concern was about the questions on topics that were seen as too personal (2/12): ``\textit{Whenever I share online, I usually want to keep a distance from my own beliefs to avoid making my comment too personal or private.}'' (P7). However, three participants found value in this reflector for deeper thinking: ``\textit{Because of this reflector, instead of formulating an opinion based on what I know right now, I start to formulate my opinion based on how my past experiences might be shaping my current understanding of issues. The reflector helps me self-evaluate myself more deeply.}'' (P1).

\paragraph{Storytelling}
Regarding the storytelling reflector, it offered participants more context and varied perspectives, akin to the persona reflector (5/12): ``\textit{There was a variety of agreements and disagreements in the stories - it helps because I was able to see and follow through different scenarios as well. That made me understand what is happening.}'' (P5).

Storytelling, while appreciated by participants suffered from the comparison with personas (4/12): ``\textit{Storytelling is engaging, less boring, and more fun. But compared to the persona, the stories are longer and less direct. The persona is shorter and more direct.}'' (P10).
Overall, it would suggest that participants, while interested in the reflector, would rather have shorter anecdotes or stories, as they did not seem to be willing to spend time engaging with the stories that much.

\paragraph{Temporal Prompts}
Regarding the last reflector, participants noted that it facilitated reflection on their personal experiences. However, they tended to rank it lower than reflectors offering alternative viewpoints (7/12): ``\textit{This reminds me to reflect on my past - the questions prompt me to contemplate my experiences, such as my gaming addiction and personal stories. Nonetheless, I rank it lower than persona and storytelling, as I prefer understanding others. Exploring various perspectives enriches my understanding, whereas this reflector solely centers on self-reflection.}''

\paragraph{Differences between reading other users' comments and Help Me Reflect}
\label{sec: users comments vs reflectors}
Participants also emphasized that the \textit{Help Me Reflect feature} offered a unique function distinct from reading other users' comments. They found it to provide a neutral and objective guidance that prompts critical thinking about an issue, as opposed to comments that might advocate a specific stance and influence agreement (9/12). One participant shared, ``\textit{A comment is already a formulated, one-sided response. It might encourage deeper thinking, but it's just another perspective to consider. The reflectors are more open-ended. They don't generate a response; instead, they prompt me to explore if I can answer those questions. It has a different impact. So, the reflectors provoke and stimulate my thinking. They prompt me to consider areas I might have overlooked in my opinion, unlike reading comments that already present a response. Comments are often just one-liners without evidence, whereas the reflectors doesn't favor whether it's objective or concise, it just triggers deeper thought.}'' (P1).

\paragraph{Fostering Self-Reflection with Help Me Reflect}
Moreover, participants reported that \textit{Help Me Reflect} facilitated valuable self-reflection on the discussion topic and themselves (12/12). Additionally, these reflectors aided in the construction of their thoughts (10/12), promoted mindful commenting in online discussions (9/12), and fostered the generation of new knowledge on the topic while also encouraging consideration of the issue from various dimensions (11/12).

\subsubsection{Summary}
We observed a trend from both the qualitative and quantitative results where some reflectors were generally preferred. As a rule of thumb, participants seemed to prefer reflectors with shorter prompts or text, as well as less abstract ones. The \textit{Persona} reflector was overall the preferred one, as it offers a variety of viewpoints very quickly with short descriptions. Participants also appreciated the \textit{Temporal Prompt} as a way to reflect about their own attitude towards a problem, enabling some introspection. Finally, participants also liked \textit{Storytelling}, as a more in-depth and verbose \emph{Persona}. The lowest preference of this latter reflector compared to persona suggests that while both reflectors help participants consider other perspectives, they would rather work with shorter, bite-sized prompts. The last two reflectors scored significantly worse and tended to receive negative feedback: \textit{Cultural Prompts} was generally not well-ranked as participants found it hard to relate to, or were concerned about the intrusive nature of some questions (and thus potentially oversharing), while \textit{Analogy and Metaphors} suffered from the extra level of abstraction added by the metaphors and their imperfect mapping to the problem discussed. Because of the concerns of the participants, we excluded these two reflectors from study 2.

\subsection{External Evaluation of Textual Prompts in Study 1}
\label{sec: external evaluation study 1}
While the primary objective of study 1 is to identify the preferred and fitting reflectors for integration into online deliberation platforms, the rankings of these reflectors could be influenced by their inherent quality (i.e., the quality of the textual prompts for each reflector that is generated by the LLM). Therefore, this section is dedicated to assessing and evaluating the quality of the textual prompts across the five reflectors.

\subsubsection{Evaluative Criteria and Procedure}
When evaluating the quality of textual prompts, we draw insights from studies that assess the quality of reflective nudges. Notably, prevailing literature predominantly focuses on evaluating the written output of reflection~\cite{desjarlais2011comparative, dyment2011assessing, fines2014assessing, koole2011factors, wong1995assessing} rather than assessing the factors enabling the reflection process. This leaves a void in the evaluation of the quality of reflective nudges that facilitate the reflection process. Two prior studies~\cite{horbach2020linguistic, king1994inquiry} have examined the linguistic and reflective qualities of short question prompts. Horbach et al.~\cite{horbach2020linguistic} assessed prompts for reading comprehension, incorporating measures such as fluency, ambiguity, answerability and relevancy to evaluate the adequacy of each prompt. Additionally, King~\cite{king1994inquiry} examined techniques for provoking reflective thinking through ``question stems'', demonstrating their efficacy in developing higher-order critical thinking of various perspectives and solutions~\cite{king1994inquiry, ullmann2011architecture, lampert2006enhancing}. Our evaluation thus leverages on King's~\cite{king1994inquiry} table of question stems (see Appendix Table~\ref{tab: reflective category table}) as a criterion, aiming to assess whether our textual prompts align with these stems to promote self-reflection effectively. In our present evaluation, we incorporated criteria from these two studies to assess the quality of the textual prompts. Our evaluation employs a seven-item rubric, encompassing the initial six hierarchical criteria from Horbach et al.~\cite{horbach2020linguistic} to evaluate the linguistic qualities of the prompts and the last criterion from King~\cite{king1994inquiry} to ascertain the reflective nature of our textual prompts.

The evaluation of the quality for all 30 textual prompts (i.e., six textual prompts were generated for each of the five reflectors) involved two external human evaluators, both PhD students. Using the seven-item rubric, evaluators assessed the linguistic and reflective quality of the textual prompts. Criteria were presented to evaluators in the order depicted in Table~\ref{tab: external evaluation criteria}, with responses recorded as binary (0 for ``no'' and 1 for ``yes'') across the seven quality dimensions. Specifically, the first six rubric items followed a hierarchical structure, rendering remaining items as ``not applicable'' if certain criteria received a ``no'' response. These criteria, bolded in Table~\ref{tab: external evaluation criteria}, prevent the distortion of rubric criteria distributions for unratable prompts and help reduce the annotation workload for the evaluators~\cite{moore2022assessing, horbach2020linguistic}. Table~\ref{tab: external evaluation criteria} also presents Cohen's Kappa $\kappa$ scores as a measure of the inter-rater reliability between the two evaluators for each rubric item. Notably, \textit{Understandable} and \textit{Answerable} achieved perfect agreement, while the others are either at near-perfect or substantial agreement~\cite{viera2005understanding, mchugh2012interrater}. A high-quality textual prompt, denoting proficiency in language and reflection, was defined by a binary score of 1 for all rubric items, aligning with prior studies~\cite{horbach2020linguistic, steuer2021linguistic, king1994inquiry}.

Alongside their evaluation scores, the evaluators also provided short feedback on their rating assessments. Throughout the evaluation, the evaluators were unaware of the conditions under which the prompts were generated, ensuring impartiality (i.e., the evaluators are independent and do not have any knowledge of what the study is about).

\begin{table*}[!htbp]
\caption{The 7-item rubric used to evaluate the textual prompts. The first six rubric items follow a hierarchical structure where the bolded criteria stop the evaluation process if answered as ``no''. Bracketed numbers indicate the Cohen's Kappa $\kappa$ value for each item.}
\label{tab: external evaluation criteria}
\begin{tabular}{@{}cll@{}}
\toprule
\textbf{No.} & \textbf{Evaluative Criteria} & \textbf{Definition} \\
\midrule
1 & \begin{tabular}[c]{@{}l@{}}\textbf{Understandable} \\ ($\kappa = 1.00$) \end{tabular} & Could you understand what the textual prompt is saying/asking? \\ \hline
2 & \begin{tabular}[c]{@{}l@{}} Domain Related \\ ($\kappa = 0.621$) \end{tabular} & Is the textual prompt related to the topic of violent video games and youth violence? \\ \hline
3 & \begin{tabular}[c]{@{}l@{}}Grammatical \\ ($\kappa = 0.937$) \end{tabular} & Is the textual prompt grammatically well-formed (i.e., is it free of language errors)? \\ \hline
4 & \begin{tabular}[c]{@{}l@{}}\textbf{Clear} \\ ($\kappa = 0.874$) \end{tabular} & Is it clear what the textual prompt is saying/asking for? \\ \hline
5 & \begin{tabular}[c]{@{}l@{}}\textbf{Answerable} \\ ($\kappa = 1.00$) \end{tabular} & Are people probably able to understand/answer the textual prompt? \\ \hline
6 & \begin{tabular}[c]{@{}l@{}}Central \\ ($\kappa = 0.726$) \end{tabular} & \begin{tabular}[c]{@{}l@{}} Do you think being able to understand/answer the textual prompt is important to \\ work on the topic? \end{tabular} \\ \hline \hline
7 & \begin{tabular}[c]{@{}l@{}}Reflective \\ ($\kappa = 0.705$) \end{tabular} & \begin{tabular}[c]{@{}l@{}} Refer to the Reflective Category Table (see Appendix Table~\ref{tab: reflective category table}). Does the textual \\ prompt falls into at least one of the reflective categories? \end{tabular} \\ 
\bottomrule
\end{tabular}
\Description{A table illustrating the 7-item rubric employed for evaluating the textual prompts. They are: Understandable, Domain Related, Grammatical, Clear, Answerable, Central, and Reflective. The first six rubric items follow a hierarchical structure where the bolded criteria stop the evaluation process if answered as no. Each rubric item is also accompanied by its corresponding definition. The bracketed numbers beneath each rubric item represent the Cohen's Kappa value associated with that particular criterion.}
\end{table*}

\subsubsection{Quantitative Results}
\begin{figure*}[!htbp]
  \centering
  \includegraphics[width=1.0\textwidth]{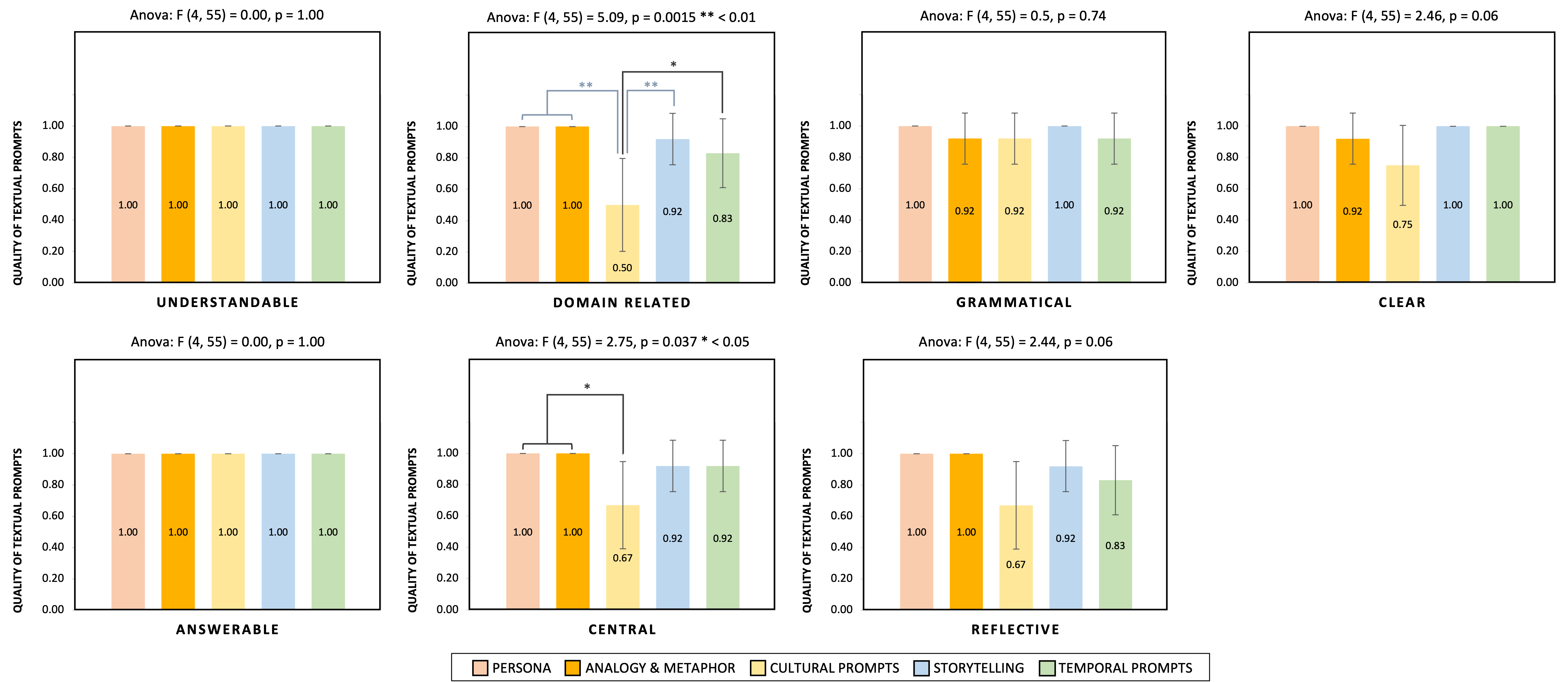}
  \caption{Mean evaluation scores on each evaluative criteria across the five reflectors for study 1. Binary assessment is applied for each evaluative criteria where 0 indicates low quality and 1 indicates high quality. Error bars show .95 confidence intervals. We report the results of the ANOVA test and pairwise comparisons with BH correction if any, where * : p < .05, ** : p < .01.}
  \label{fig: Evaluation Exp.1}
  \Description{Bar graphs describing the mean evaluation scores for the five reflectors on each evaluative criteria. The X-axis shows the seven different evaluative criteria and the Y-axis shows the evaluation scores ranging from 0 to 1 at a rank interval of 0.2. 0 is the lowest possible evaluation score and 1 is the highest possible evaluation score. Cultural prompts ranked significantly lower than the rest of the reflectors in the evaluative criteria of Domain Related and Central. No other significant main effect and pairwise differences were found between the reflectors and the evaluative dimensions.}
\end{figure*}

A summary of the evaluation scores on each of the seven evaluative criteria across the five reflectors is shown in Figure~\ref{fig: Evaluation Exp.1}. Particularly, a one-way repeated measures ANOVA was conducted to investigate the influence of the Reflectors on the seven evaluative criteria. Our findings revealed a significant main effect of the Reflectors on two evaluative criteria: Domain Related ($F_{4,55} = 5.09$, $p<.01$) and Central ($F_{4,55} = 2.75$, $p<.05$). Post-hoc pairwise t-tests revealed that in terms of Domain Related, the evaluative score for \emph{Cultural Prompts} ($M = 0.50$) is significantly lower than the rest of the reflectors: \emph{Persona} ($M = 1.00, p<.01$), \emph{Analogies and Metaphors} ($M = 1.00, p<.01$), \emph{Storytelling} ($M = 0.92, p<.01$) and \emph{Temporal Prompts} ($M = 0.83, p<.05$). Similarly, for the Central criteria, the evaluative score for \emph{Cultural Prompts} ($M = 0.67$) is also significantly lower when compared to \emph{Persona} and \emph{Analogies and Metaphors} (both $M = 1.00, p<.05$). We found no other significant differences among the evaluative scores on all other evaluative criteria among the reflectors.

\subsubsection{Qualitative Results}
\label{sec: qual results for external evaluation 1}
To gain insights into the evaluation scores, we asked evaluators to write a short feedback on their rating assessments. 

Overall, evaluators noted the high relevance of all the textual prompts to the topic, with the exception of \emph{Cultural Prompts}, and acknowledged their efficacy in prompting reflection in individuals. In the \emph{Persona} reflector, they noted the presentation of ``\textit{distinct viewpoints and perspectives}'', contributing ``\textit{potentially real-life examples for discussing the topics}''. In the case of \emph{Analogies and Metaphors}, the evaluators found the textual prompts to be ``\textit{thought-inducing and captivating, incorporating vivid metaphors on the topic}''. For \emph{Storytelling}, evaluators praised their detailed, thorough explanation, emphasizing their ``\textit{excellence in illustrating the complex influence of violent video games on youths}''. They also recognized the \emph{Storytelling} prompts for their ability to stimulate deeper reflection by ``\textit{allowing individuals to empathized with the characters, engage in critical thinking and contemplate on higher-level thoughts}''. Similarly, evaluators identified the textual prompts from \emph{Temporal Prompts} as ``\textit{thought-inducing}'' and ``\textit{effective in prompting individuals to reflect on their personal experiences, offering insightful perspectives on the relationship between video games and their lives}''. They also highlighted how these prompts ``\textit{encourage individuals to be more objective}''.

Finally, evaluators found that the textual prompts from the \emph{Cultural Prompts} reflector were ``\textit{thought-inducing}'' but not directly relevant to the topic. ``\textit{The textual prompts do not relate to violent video games or youth violence but they help encourage people to think about the power of biases, so that people would gain more insights and be less influenced by public voices or personal opinions.}'' This observation is attributed to the nature of the cultural prompts reflector, emphasizing reflection on beliefs and biases~\cite{adams2003reflexive, donati2011modernization, mouzelis2010self}, which, being distinct from the topic, contributed to lower scores in the evaluative criteria of Domain Related and Central. However, this does not indicate a lower LLM-generated prompt quality; rather, it reflects the intrinsic nature of the cultural prompt reflector.

Our findings suggest no significant differences in the quality of textual prompts across the five reflectors. Variances in quality, notably observed in \emph{Cultural Prompts}, are attributed to the inherent nature of the reflector rather than deficiencies in LLM-generated prompt quality. An improvement for future external evaluations could involve domain experts to assess textual prompt quality. However, given the limited existing research on appraising factors that facilitate reflection, identifying suitable experts poses challenges. Future evaluation work may benefit by involving experts well-versed in the domain of evaluating reflective nudges.
\section{Study 2: Deliberativeness of the Top 3 Reflectors}
From study 1, we found the significantly preferred reflectors were \textit{Persona}, \textit{Storytelling} and \textit{Temporal Prompts}. For study 2, we thus compared each reflector to a control condition with no reflector to find out how the reflectors may improve deliberativeness.

\subsection{Independent Variables and Experimental Design}
A between-subject experiment was conducted with the independent variable being the Reflector with four levels: \{Control, Persona, Storytelling, Temporal prompts\}. The experimental interface retained the same design as that used in study 1 (refer to Figure~\ref{fig: features}), but for this study, the \textit{Help Me Reflect} feature was limited to a single reflector. Participants in the control group would see the same interface without the \textit{Help Me Reflect} button.

\subsection{Dependent Variables}
As deliberativeness is multi-dimensional, we operationalized it through five measurements: argument repertoire, argument diversity, rationality (opinion expression), rationality (justification level) and constructiveness as discussed in section~\ref{sec: measurements}. 

All five metrics were derived from a content analysis of participants' responses by two coders. Both coders were PhD students with respectively 2 and 5 years of experience using content analysis. Cohen's Kappa was used to determine the agreement between the two coders’ judgments: Argument Repertoire ($\kappa = 0.739$); Argument Diversity ($\kappa = 0.739$); Rationality (Opinion) ($\kappa = 0.790$); Rationality (Justification Level) ($\kappa = 0.859$); Constructiveness ($\kappa = 0.783$). Kappa scores for all metrics were above the satisfactory threshold of 0.70~\cite{viera2005understanding, mchugh2012interrater}. 

The dependent variables are coded as follows:
\begin{itemize}
    \item \textit{Argument repertoire} is the number of non-redundant arguments regarding each position of the discussion topic. The ideas produced along the two positions were combined.
    \item \textit{Argument diversity} was coded by counting the number of unique themes present in the entire response. A higher diversity count indicates more varied perspectives present in the participant's responses~\cite{anderson2016all, gao2023coaicoder}.   
    \item \textit{Rationality (opinion)} captures whether opinions are expressed or information is provided in the response. This was coded with two levels: 1) no opinions were expressed, rather, information was provided (score of 0); 2) an opinion, personal assertion or a claim was made (score of 1). This also includes evaluation, a personal judgment or assertion. 
    \item \textit{Rationality (justification level)} captures the degree to which reasons are used to justify one's claims. This were coded at four levels: 1) no justification was provided (score of 0); 2) an inferior justification was made - this indicates that the opinion is supported with a reason in an associational way such as through personal experiences or an incomplete inference was given (score of 1); 3) a qualified justification was made when there is a single complete inference provided in the opinion (score of 2); 4) a sophisticated justification was made when at least two complete inferences was provided (score of 3).  
    \item \textit{Constructiveness} captures the degree of balance within an opinion. This was coded at two levels: 1) the opinion is one-sided (score of 0); 2) the opinion is two-sided when multiple perspectives and viewpoints are presented (score of 1). 
\end{itemize}

\subsection{Power Analysis}
 We conducted a power calculation for a four-group ANOVA study seeking a medium effect size (0.30) according to Cohen’s conventions, at 0.80 observed power with an alpha of 0.05, giving $N=30$ per experimental condition, hence we recruited 120 participants. 

\subsection{Participants and Ethics}
A total of 120 participants were recruited through Amazon Mechanical Turk. Refer to Appendix Table~\ref{tab: demograhics study 2} on the breakdown of the demographic profile in each experimental condition. We ensured that the demographic profiles across the four conditions were similar so as to control for any fixed effects resulting from the differences in demographic factors. Similar to study 1, we got ethics approval from our local IRB and reimbursed participants at an hourly rate of 7.25 USD.

\subsection{Procedure and Task}
The procedure for this study closely follows that of study 1 outlined in section~\ref{procedure}. In the pre-task phase, we gathered data on demographics and provided instructions on reflection.

In the main task, participants were instructed to utilize the \textit{Help Me Reflect} feature, which exclusively presents textual prompts from a single reflector (either persona, storytelling, or temporal prompts). As we focus on textual prompts from a specific reflector, we instructed the LLM to generate an additional six distinct textual prompts for each of the three reflectors, resulting in a total of 12 textual prompts per reflector. Similar to study 1, the topic remained the same, requiring participants to type and post a response on the problem of video games and violence. We maintain consistency in the topic across the two studies to draw robust comparisons between the different reflectors and their impact on deliberativeness within the same thematic domain~\cite{cahit2015internal}.

In the post-task phase, participants completed a short survey in which they could provide feedback on the reflector they engaged with.

\subsection{Outliers}
Before analyzing our data, we plotted a box plot (box and whisker plot) to visually show the dispersion of our data and to identify any potential outliers~\cite{schwertman2004simple}. We observed three abnormalities in the data and adopted the approach by Dawson~\cite{dawson2011significant} to discard data points that were either below $Q1 - 3 IQR$ or above $Q3 + 3 IQR$ (where Q1 is the first quartile, Q3 the third, and IQR is the interquartile range). This included one data point from the control group and two data points from storytelling. Hence, our final sample size was 117.  
\section{Results (Study 2)}
 
\subsection{Quantitative Results}
\label{sec: quantitative}
Unless otherwise specified, ANCOVA was used to identify main effects while controlling for covariates (i.e. Demographics, TK-TI and SRIS), and pairwise t-tests with Benjamini-Hochberg correction for post-hoc comparisons for the five measures of opinion quality. We did not use Bonferroni correction, as its conservative approach leads to high rates of false negatives when done with large number of comparisons~\cite{thissen2002quick}. Instead, we relied on Benjamini-Hochberg as it minimizes the problem~\cite{nakagawa2004farewell} while still accounting for multiple comparisons. We report effects of the covariates only where they are significant (namely for Argument Repertoire).

\subsubsection{Argument Repertoire}
\begin{figure*}[!htbp]
  \centering
  \includegraphics[width=.6\textwidth]{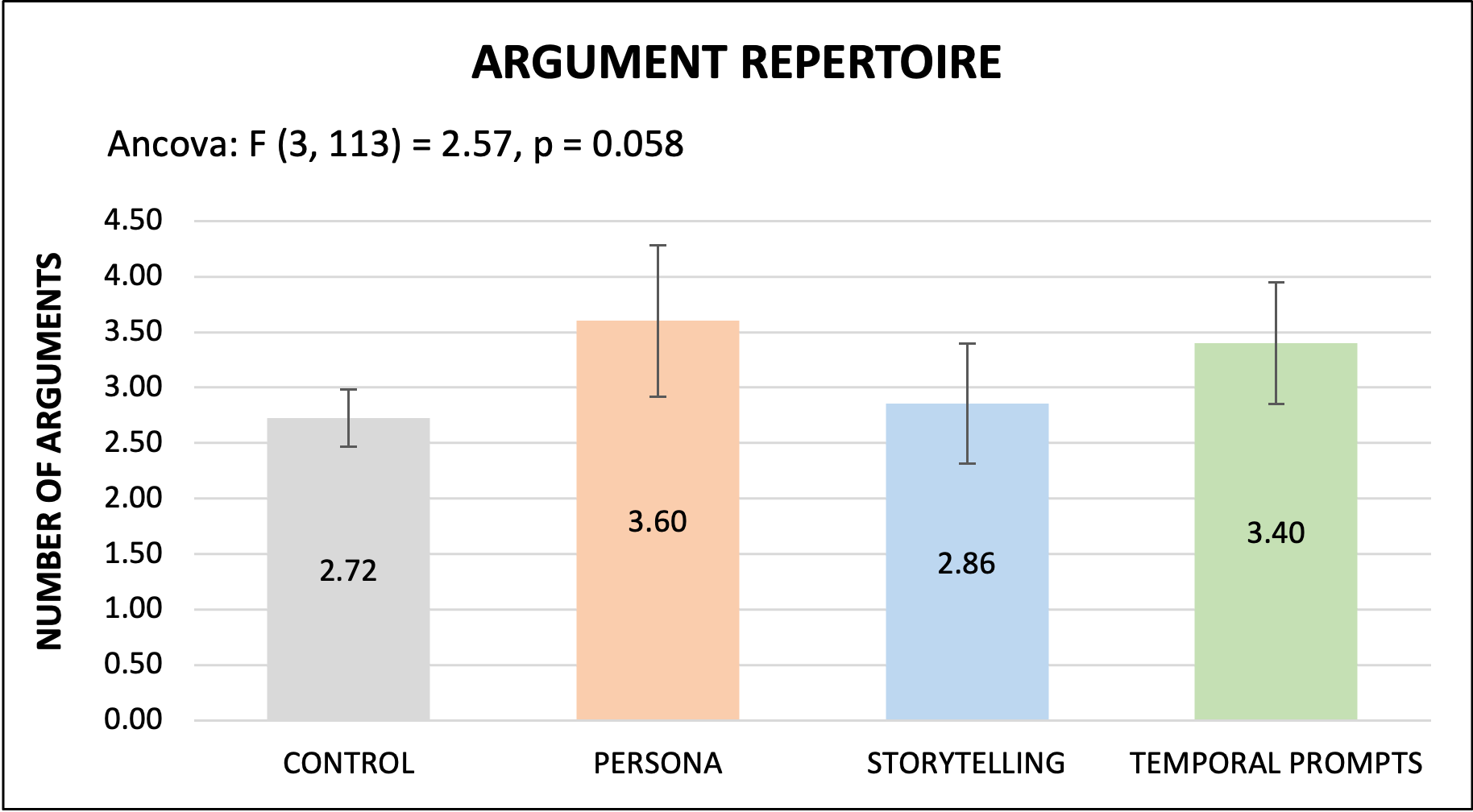}
  \caption{Argument Repertoire. Error bars show .95 confidence intervals. We report the results of the ANCOVA test, and pairwise comparisons with BH correction if any, where * : p < .05, ** : p < .01.}
  \label{fig: argument repertoire}
  \Description{Bar graph describing the results for argument repertoire. The X-axis shows the four conditions: control, persona, storytelling and temporal prompts, and the Y-axis shows the number of arguments ranging from 0 to 4.50 at an interval of 0.50. Opinions in the persona condition has the highest argument repertoire at 3.60, followed by temporal prompts at 3.40 and storytelling at 2.86. The control group has the lowest argument repertoire at 2.72.}
\end{figure*}

We found no statistically significant main effect of the Reflectors on Argument Repertoire after controlling for TK-TI and SRIS ($F_{3,113} = 2.57$, $p = 0.058$), although having a higher topic knowledge and interest ($p<.05$) and tendency to self-reflect ($p<.01$) are associated with higher argument repertoire. The results are summarized in Figure~\ref{fig: argument repertoire}. 

\subsubsection{Argument Diversity}
\begin{figure*}[!htbp]
  \centering
  \includegraphics[width=.6\textwidth]{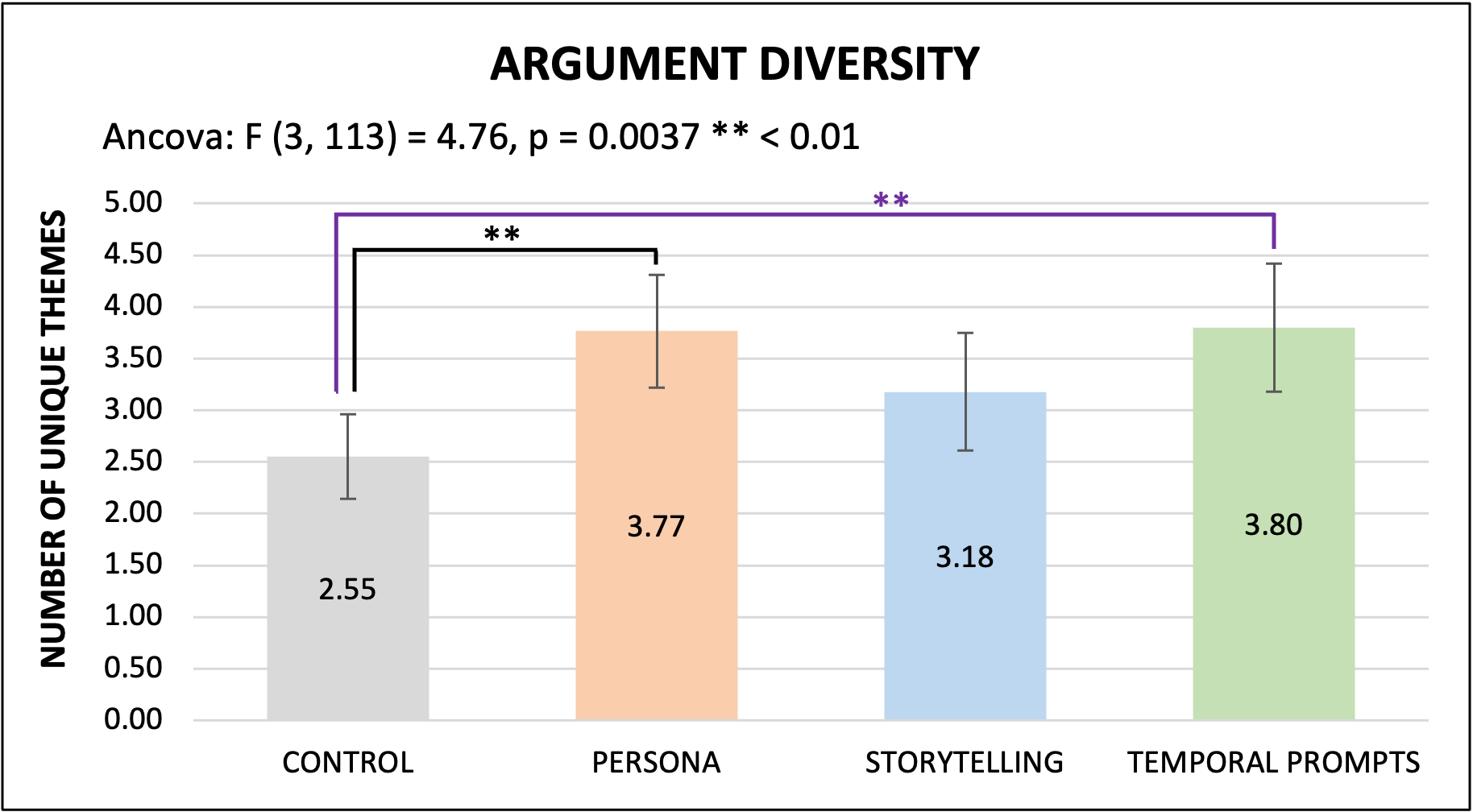}
  \caption{Argument Diversity. Error bars show .95 confidence intervals. We report the results of the ANCOVA test, and pairwise comparisons with BH correction if any, where * : p < .05, ** : p < .01. Note: a single argument may have multiples themes.}
  \label{fig: argument diversity}
  \Description{Bar graph describing the results for argument diversity. The X-axis shows the four conditions: control, persona, storytelling and temporal prompts, and the Y-axis shows the number of unique themes ranging from 0 to 5.00 at an interval of 0.50. Opinions in the temporal prompts condition has the highest argument diversity at 3.80, followed by persona at 3.77 and storytelling at 3.18. The control group has the lowest argument diversity at 2.55. Note: a single argument may have multiples themes.}
\end{figure*}

We found a significant main effect of the Reflectors on Argument Diversity ($F_{3,113} = 4.76$, $p<.01$). There was a statistical difference between \emph{Control} ($M = 2.55$ themes) and \emph{Persona} ($M = 3.77$ themes, $p<.01$). We also found a significant difference between \emph{Control} and \emph{Temporal Prompts} ($M = 3.80$ themes, $p<.01$). The results are summarized in Figure \ref{fig: argument diversity}.

\subsubsection{Rationality (Opinion)}
\begin{figure*}[!htbp]
  \centering
  \includegraphics[width=.6\textwidth]{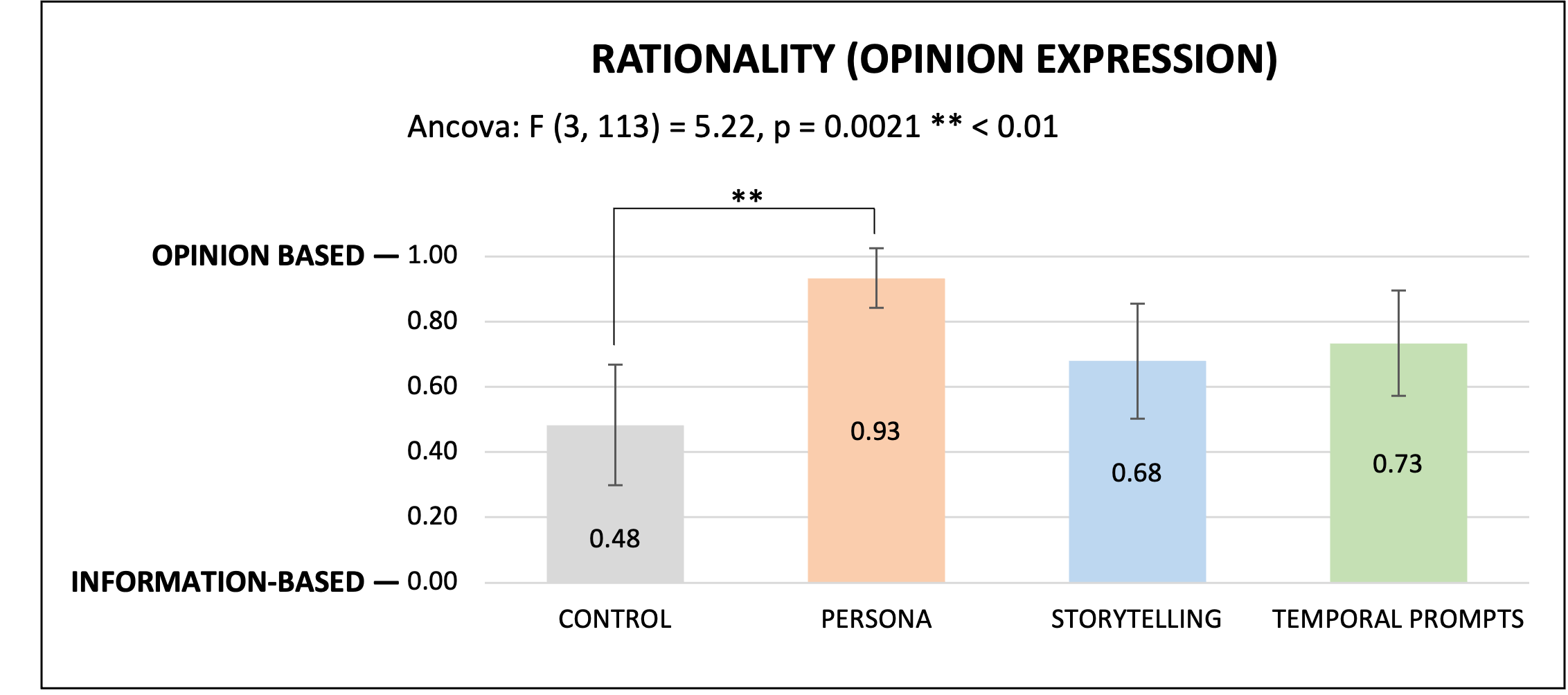}
  \caption{Rationality (Opinion) which is coded on two levels: 0 indicates that the opinions lean towards more information-based and 1 indicates that the opinions are more opinionated and discussion-oriented. Error bars show .95 confidence intervals. We report the results of the ANCOVA test, and pairwise comparisons with BH correction if any, where * : p < .05, ** : p < .01.}
  \label{fig: rationality opinion expression}
  \Description{Bar graph describing the results for rationality (opinion). The X-axis shows the four conditions: control, persona, storytelling and temporal prompts, and the Y-axis shows opinion expression ranging from 0 to 1.00. Opinion expression is coded on two levels: 0 indicates that the opinions lean towards more information-based and 1 indicates that the opinions are more opinionated and discussion-oriented. Participants in the persona condition have the highest opinion expression at 0.93, followed by temporal prompts at 0.73 and storytelling at 0.68. The control group has the lowest opinion expression at 0.48.}
\end{figure*}

We found a significant main effect of the Reflectors on the Opinion dependent variable ($F_{3,113} = 5.22$, $p<.01$). There was a statistical difference between \emph{Control} ($M = 0.48$) and \emph{Persona} ($M = 0.93$, $p<.01$). We did not find any other pairwise differences. The results are summarized in Figure \ref{fig: rationality opinion expression}. 

\subsubsection{Rationality (Justification Level)}
\begin{figure*}[!htbp]
  \centering
  \includegraphics[width=.6\textwidth]{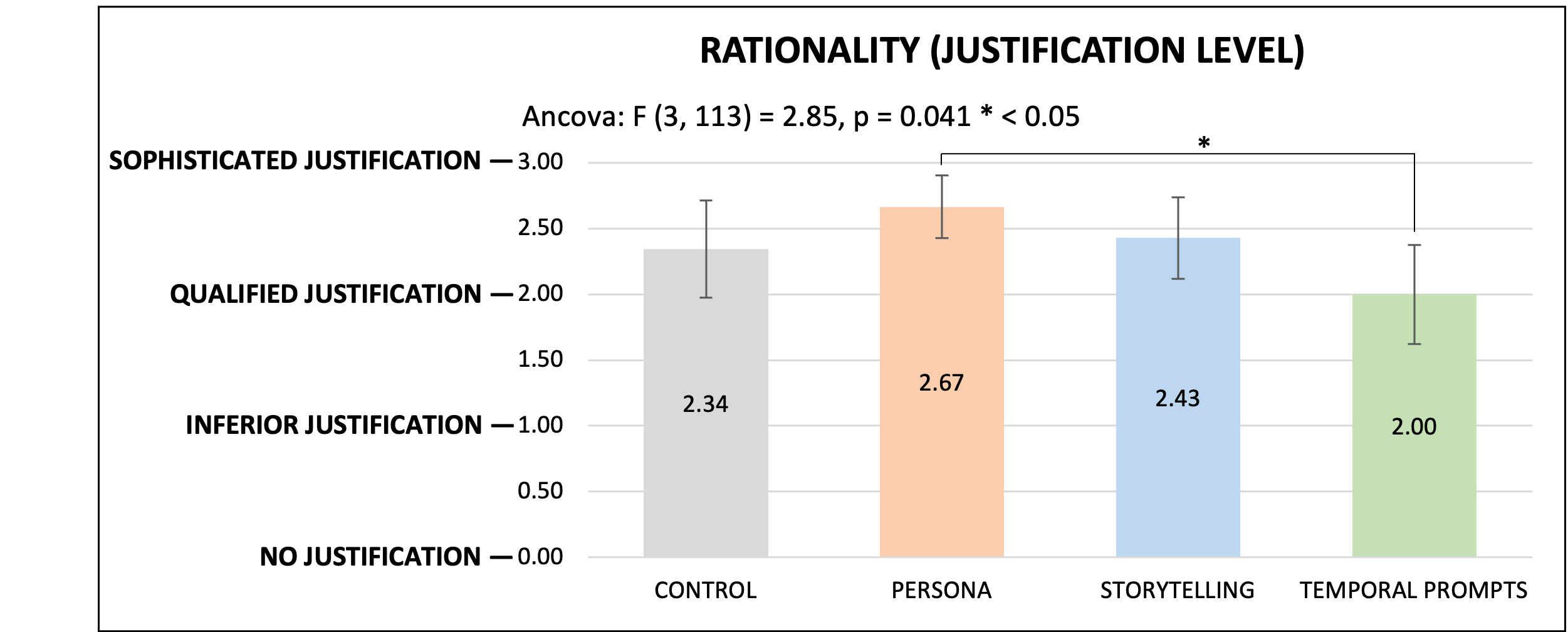}
  \caption{Rationality (Justification Level) which is coded on four levels, ranging from the absence of justifications to the presence of strong valid reasonings: 0 for no justification, 1 for inferior justification, 2 for qualified justification and 3 for sophisticated justification. Error bars show .95 confidence intervals. We report the results of the ANCOVA test, and pairwise comparisons with BH correction if any, where * : p < .05, ** : p < .01.}
  \label{fig: rationality justification level}
  \Description{Bar graph describing the results for rationality (justification level). The X-axis shows the four conditions: control, persona, storytelling and temporal prompts, and the Y-axis shows the level of justification ranging from 0 to 3.00. Justification level is coded on four levels, ranging from the absence of justifications to the presence of strong valid reasonings: 0 for no justification, 1 for inferior justification, 2 for qualified justification and 3 for sophisticated justification. Opinions in the persona condition have the highest justification level at 2.67, followed by storytelling at 2.43 and the control group at 2.34. Temporal prompts has the lowest justification level at 2.00.}
\end{figure*}

We found a significant main effect of the Reflectors on Justification Level ($F_{3,113} = 2.85$, $p<.05$). There was a statistical difference between \emph{Persona} ($M = 2.67$) and \emph{Temporal Prompts} ($M = 2.00$, $p<.05$). We did not find any other pairwise differences. The results are summarized in Figure \ref{fig: rationality justification level}. 

\subsubsection{Constructiveness}
\begin{figure*}[!htbp]
  \centering
  \includegraphics[width=.6\textwidth]{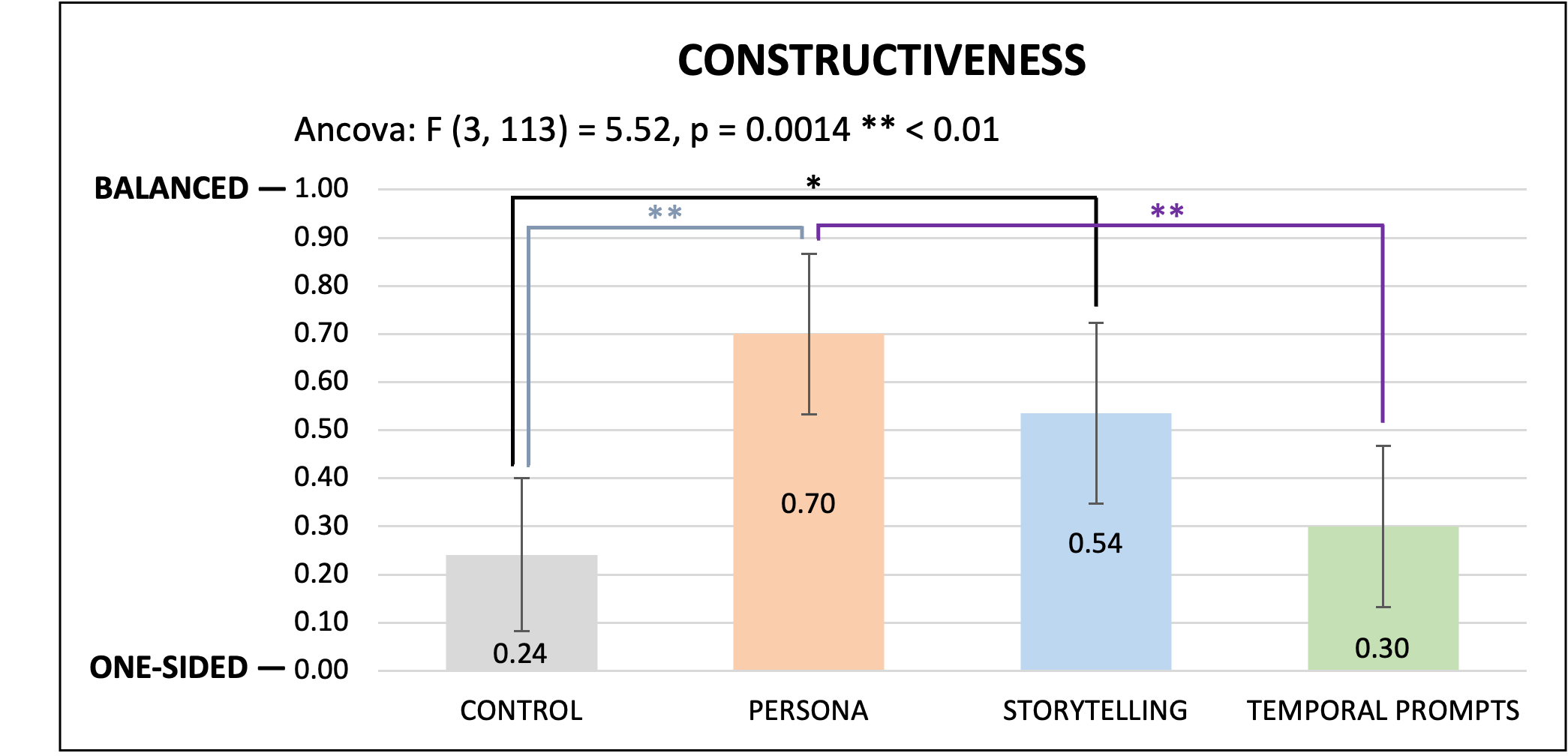}
  \caption{Constructiveness which is coded on two levels: 0 indicates that the opinions are one-sided and 1 indicates that the opinions are balanced. We report the results of the ANCOVA test, and pairwise comparisons with BH correction if any, where * : p < .05, ** : p < .01.}
  \label{fig: constructiveness}
  \Description{Bar graph describing the results for constructiveness. The X-axis shows the four conditions: control, persona, storytelling and temporal prompts, and the Y-axis shows the degree of constructiveness from 0 to 1.00. Constructiveness is coded on two levels: 0 indicates that the opinions are one-sided and 1 indicates that the opinions are balanced. Opinions in the persona condition have the highest constructiveness at 0.70, followed by storytelling at 0.54 and temporal prompts at 0.30. Opinions from the control group have the lowest constructiveness at 0.24.}
\end{figure*}

We found a significant main effect of the Reflectors on Constructiveness ($F_{3,113}$ = 5.52, $p<.01$). There was a significant difference between \emph{Control} ($M = 0.24$) and \emph{Persona} ($M = 0.70$, $p<.01$), as well as between \emph{Control} and \emph{Storytelling} ($M = 0.54$, $p<.05$). Lastly, we also found a difference between \emph{Persona} and \emph{Temporal Prompts} ($M = 0.30$, $p<.01$). The results are summarized in Figure \ref{fig: constructiveness}. 

\subsubsection{Summary of Results} 
Overall, the deliberative quality by the five measurements is summarized in Table~\ref{tab: summary quantitative}. A careful look at the data reveals that \emph{Persona} achieves the highest overall opinion quality in terms of Diversity, Opinion Expression, Justification Level and Constructiveness, in most cases significantly higher than Control, suggesting it to be the best reflector, enabling users to embrace multiple viewpoints. \emph{Storytelling} only outperforms \emph{Control} in terms of Constructiveness, while \emph{Temporal Prompts} leads to higher Diversity at the expense of Justification Level, showing that while its presence may increase deliberativeness, the contribution are still mostly one-sided and focused on the user's opinion. 

\begin{table*}[!htbp]
\caption{Deliberative quality across the different reflectors and control group}
\label{tab: summary quantitative}
\begin{tabular}{@{}ccccc@{}}
\toprule
\multirow{3}{*}{\textbf{Deliberativeness}} & \multicolumn{4}{c}{\textbf{Conditions}} \\ \cline{2-5}
& \textbf{\begin{tabular}[c]{@{}c@{}}Control \\ ($M \pm S.D.$) \end{tabular}} & \textbf{\begin{tabular}[c]{@{}c@{}} Persona \\ ($M \pm S.D.$) \end{tabular}} & \textbf{\begin{tabular}[c]{@{}c@{}} Storytelling \\ ($M \pm S.D.$) \end{tabular}} & \textbf{\begin{tabular}[c]{@{}c@{}} Temporal Prompts \\ ($M \pm S.D.$) \end{tabular}} \\
\midrule
Argument Repertoire & $2.72 \pm 0.70$ & $3.60 \pm 1.90$ & $2.86 \pm 1.46$ & $3.40 \pm 1.52$ \\
Argument Diversity & $2.55 \pm 1.12$ & $3.77 \pm 1.52$ & $3.18 \pm 1.54$ & $3.80 \pm 1.73$ \\
Rationality (Opinion Expression) & $0.48 \pm 0.51$ & $0.93 \pm 0.25$ & $0.68 \pm 0.48$ & $0.73 \pm 0.45$ \\
Rationality (Justification Level) & $2.34 \pm 1.01$ & $2.67 \pm 0.66$ & $2.43 \pm 0.84$ & $2.00 \pm 1.05$ \\
Constructiveness & $0.24 \pm 0.44$ & $0.70 \pm 0.47$ & $0.54 \pm 0.51$ & $0.30 \pm 0.47$ \\
\bottomrule
\end{tabular}
\Description{A table summarizing the deliberative quality across the four conditions: control, persona, storytelling and temporal prompts, and across the dependent variables. The table is a summary of the other Figures in the section.}
\end{table*}

\subsection{Qualitative Results}
\label{sec: qualitative}
In this section, we present our findings derived from the content analysis of participants' opinions, as well as their post-task feedback. Numbers enclosed in parentheses represent the frequency of occurrence of specific patterns within each condition. 
\subsubsection{Persona: Identifying Factors contributing to Violence in Video Games}
Opinions expressed within the \emph{Persona} group exhibited a higher degree of evaluation (11) compared to both the control (2) and storytelling (2) groups ($p<.01$). Specifically, participants within the persona group consistently emphasized that attributing violence solely to video games is overly simplistic (P4, 8, 17, 23), highlighting the presence of various contributing factors like parental influence and social environment (P2, 9, 12, 22, 29). For instance, a participant from the persona group remarked, ``\textit{The impact of violent video games on youth behavior is not a straightforward matter. Scientific studies do not unequivocally establish a direct link between such games and aggression. If a child is influenced by them, it might stem from issues with parenting, mental health concerns, bullying, and other contextual factors}'' (P2). This underscores the capacity of the persona reflector to facilitate a broader perspective among participants.

Feedback from the post-task is aligned with the qualitative analysis, where the main reported advantage of \emph{Persona} was the ability to cover multiple viewpoints (16): ``\textit{It helps me to see different points of views on both sides of the issue/question. I get to understand why they think so and why they didn't. It opens up my mind and helps me to be more open-minded. I can change my mind after reading it if it seems reasonable and if it's what I haven't thought of}'' (P6). See Table~\ref{tab: top 3 post-task feedback} for the post-task feedback on persona.

\subsubsection{Persona and Temporal Prompts: Voicing Opinions}
Furthermore, we also noticed that in both experimental conditions: persona and temporal prompts, participants engaging with these conditions displayed a greater inclination to voice their opinions and clearly articulate their stance on the given issue. Expression of one's opinions were significantly higher in persona (24) ($p<.01$) and temporal prompts (21) ($p<.05$) groups as compared to the control group (11). This observation is in line with the quantitative findings related to rationality of opinions. Specifically, the control group predominantly provides information based responses, whereas the persona group exhibited a proclivity for opinion expression.

\subsubsection{Temporal Prompts: Sharing Personal Experiences}
Another observation pertains to the experimental group engaged with temporal prompts. This group exhibited a distinct inclination toward incorporating personal experiences within their opinions. Notably, the presence of personal experiences in their opinions (10) was significantly more pronounced in comparison to the control group (1) ($p<.01$), persona group (4) ($p<.05$), and storytelling group (1) ($p<.01$). This observation can be attributed to the inherent nature of the temporal prompts, which explicitly encourage users to reflect on their individual life journeys and personal experiences. This phenomenon could potentially explain our quantitative findings concerning constructiveness and rationality (justification level). This may explain the lower degree of constructiveness observed in this condition due to a propensity for one-sidedness.

\subsubsection{Temporal Prompts, Post-task feedback: Heightened Thinking}
The primary positive aspect reported in the post-task questionnaire revolves around heightened thinking (9) and knowledge acquisition (5). Participants expressed their sentiments, saying ``\textit{I enjoy how Help Me Reflect inspires us to think about the issues in our own way}'' (P28) and ``\textit{I like how it helped me to stop and take time to think about how the topic may have affected me in the past}'' (P22). In knowledge acquisition, participants commented that they gained more knowledge (P5, 16) and that increase their learning of the topic (P5, 7): ``\textit{It will help you achieve a new higher level of consciousness and it may just help you find valuable information}'' (P3). As participants engaged in reflective thinking about their own personal experiences, this heightened level of thinking likely contributed to the observed high degree of argument diversity within the temporal prompt condition. See Table~\ref{tab: top 3 post-task feedback} for the post-task feedback on temporal prompts.

\subsubsection{Storytelling, Post-task feedback: Knowledge Acquisition}
For the \emph{storytelling} condition, a key benefit is the acquisition of knowledge (7). Participants emphasized this aspect by stating that the reflector was helpful in providing them knowledge (P24, 27) and additional insights (P14) through multiple stories. This in turn was helpful for participants to learn more about the topic (P16, 21) and they find it useful as they gained lots of information (P18, 21). These benefits were succinctly described in P15 feedback: ``\textit{I learned about another aspect of gaming which, while I was aware of it, was not aware of the extent to which it could benefit a person.}'' See Table~\ref{tab: top 3 post-task feedback} for the post-task feedback on storytelling.

\subsubsection{Control: Users tended to not identify Positive Effects}
Lastly, we noticed that in the control group, participants exhibited a tendency to be less prone to highlighting positive effects in their responses. Instead, they appeared more aligned with the notion of the question that violent video games are implicated in youth violence. Conversely, all other experimental conditions demonstrated a heightened inclination to acknowledge the presence of positive effects. Notably, this trend in the control group (2) was significantly lower when compared to the storytelling (10) and temporal prompts (11) conditions ($p<.05$). 

These observations collectively suggest that the mere presence of the \textit{Help Me Reflect} feature, along with the incorporation of different reflectors, significantly influenced participants' perceptions of the topic, encouraging them to delve into both positive and negative dimensions of the issue in a more nuanced manner. 

\subsubsection{Post-task: Opinion on the System}
\label{post-task 2}

Among the top three positive aspects across these reflectors was participants' appreciation for the \textit{Help Me Reflect} feature (see Table~\ref{tab: top 3 post-task feedback}). For the \emph{Persona} reflector, participants found the feature to be useful for reflection (10): ``\textit{I liked the overall functionality of the feature as it provided different perspectives on the topic. I specifically liked that it gave the age of the person. This is useful for me because it provided the context of both the person and the topic}'' (P19). Similarly, in the case of \emph{Storytelling}, participants recognized the value of this feature (17) with comments such as ``\textit{I appreciated the format of each story, presented in digestible pages, enabling comprehensive comprehension at one's own pace}'' (P9). Within the \emph{Temporal Prompts} condition, users found the feature well-suited for discussion platforms like Reddit (18), with P2 mentioning ``\textit{Help Me Reflect is very relevant and it is a good fit for this platform like Reddit. There's nothing I dislike about it.}'' Additional feedback highlighted the user-friendliness, straightforwardness, and ease of use of the feature. 

\begin{table*}[!htbp]
\caption{Top Three Post-Task Feedback for Each Reflector: The numbers represent the frequency of occurrence of specific feedback mentioned by participants for each condition. The feedback are listed in descending order, from the highest to the lowest.}
\label{tab: top 3 post-task feedback}
\begin{tabular}{@{}ccccc@{}}
\toprule
\textbf{Reflectors} & \multicolumn{2}{c}{\textbf{Positives}} & \multicolumn{2}{c}{\textbf{Negatives}} \\ 
\midrule
\multirow{3}{*}{\textbf{Persona}} & Different viewpoints & 16 & Fact-check information & 4 \\
& Good interface & 10 & More explanation & 3 \\
& Heightened thinking & 4 & Increase font size & 3 \\
\hline
\multirow{3}{*}{\textbf{Storytelling}} & Good interface & 17 & Improvements for interface & 8 \\
& Knowledge acquisition & 7 & More explanation & 3 \\
& Good stories & 6 & Increase font size & 1 \\
\hline
\multirow{3}{*}{\textbf{Temporal Prompts}} & Good interface & 18 & Increase font size & 3 \\
& Heightened thinking & 9 & Improvements for interface & 3 \\
& Knowledge acquisition & 5 & More explanation & 1 \\
\bottomrule
\end{tabular}
\Description{A table summarizing the top three post-task feedback for each reflector: persona, storytelling and temporal prompts.}
\end{table*}

For a comprehensive review of the positive and negative feedback associated with the three experimental conditions, please refer to Appendix Figures~\ref{fig: persona post-task}, ~\ref{fig: storytelling post-task}, ~\ref{fig: temporal prompts post-task} and Appendix Tables~\ref{tab: pesona post-task feedback}, ~\ref{tab: storytelling post-task feedback}, ~\ref{tab: temporal prompts post-task feedback}. Appendix Figure~\ref{fig: summary of results study 2} also provides a visual summary of both quantitative and qualitative results. 

\subsection{External Evaluation of Textual Prompts in Study 2}
Similar to section~\ref{sec: external evaluation study 1}, we evaluate the quality of the textual prompts for \emph{Persona}, \emph{Storytelling} and \emph{Temporal Prompts}. The evaluation of the quality for all 36 textual prompts (i.e., 12 textual prompts were generated for each of the three reflectors) involved the same two external evaluators. Likewise, using the same seven-item rubric, the evaluators assessed the linguistic and reflective quality of the textual prompts. Feedback were also provided for the additional six textual prompts that were generated by the LLM for each of the reflector.

\subsubsection{Quantitative Results}
\begin{figure*}[!htbp]
  \centering
  \includegraphics[width=1.0\textwidth]{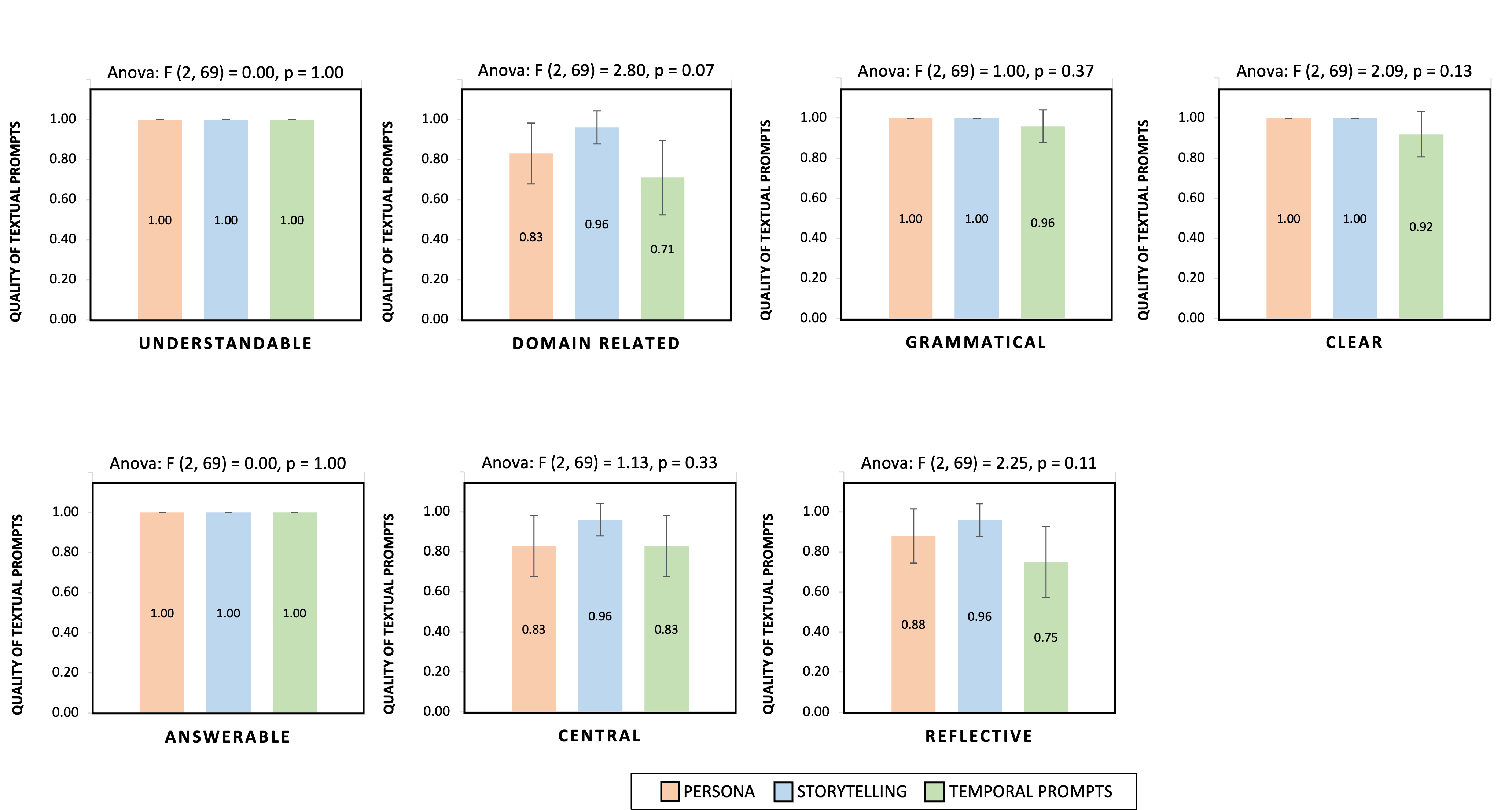}
  \caption{Mean evaluation scores on each evaluative criteria across the three reflectors for study 2. Binary assessment is applied for each evaluative criteria where 0 indicates low quality and 1 indicates high quality. Error bars show .95 confidence intervals. We report the results of the ANOVA test and pairwise comparisons with BH correction if any, where * : p < .05, ** : p < .01.}
  \label{fig: Evaluation Exp.2}
  \Description{Bar graph describing the mean evaluation scores for the three reflectors on each evaluative criteria. The X-axis shows the seven different evaluative criteria and the Y-axis shows the evaluation scores ranging from 0 to 1 at a rank interval of 0.2. 0 is the lowest possible evaluation score and 1 is the highest possible evaluation score. No significant main effect and pairwise differences were found across the three reflectors and the evaluative dimensions.}
\end{figure*}

A summary of the evaluation scores on each of the seven evaluative criteria across the three reflectors is shown in Figure~\ref{fig: Evaluation Exp.2}. A one-way repeated measures ANOVA was conducted to investigate the influence of the Reflectors on the seven evaluative criteria. Our findings revealed no significant main effect of the Reflectors on all the seven evaluative criteria. No significant differences in the quality of the textual prompts for all seven evaluative criteria were also found between any of the three reflectors.

\subsubsection{Qualitative Results}
Evaluators consistently highlighted the high clarity and relevance of the textual prompts, emphasizing their capacity to stimulate reflective thinking. Similar to the qualitative results reported in section~\ref{sec: qual results for external evaluation 1}, the textual prompts from \emph{Persona} were commended for their ``\textit{potentially real-life examples to demonstrate unique viewpoints on the influences of violent video games}''. \emph{Storytelling} were lauded as ``\textit{good stories that are not so straightforward, thus stimulating critical thinking and deep self-reflection}''. For \emph{Temporal Prompts}, evaluators mentioned that the textual prompts are``\textit{very good questions that could help individuals reflect on the relationship between video games and violence, which is crucial for the topic discussion}''. 
\section{Discussion}
In this section, we discuss our findings to answer \textbf{RQ2}. We synthesized both quantitative data from our objective measures (see section~\ref{sec: quantitative}) and qualitative data from our subjective feedback (see section~\ref{sec: qualitative}) as well as findings from our initial study to provide a more holistic view of the impacts of the reflectors on deliberativeness. We also discussed how our results corroborate and enrich previous studies as well as the implications of increasing deliberativeness on online deliberation platforms.

\subsection{Supporting Deliberation with Reflectors}
\subsubsection{Reflectors serve as catalyst for self-reflection} 
Our findings indicate that the reflectors have a positive impact on users' self-reflection and their ability to compose better opinions before posting online. Although argument repertoire was similar across all the conditions, distinctions emerged in terms of the quality of the arguments. Similarly, it is also worth noting that topic knowledge and interest (TK-TI) and reflection level (SRIS) were associated with higher argument repertoire, but did not appear to influence the quality of other deliberative dimensions.

The persona reflector provided the highest deliberative quality. The ``imagined other'' approach~\cite{chang2008personas, zhang2021nudge} of reflection not only exposed participants to different viewpoints but also appealed to human empathy~\cite{muradova2021seeing}, helping participants to be more open-minded and see the issue in a different light. For the storytelling reflector, participants reported that they acquired knowledge on the discussion topic simply by perusing various short stories which enhanced their understanding and offered extra insights into the topic, resonating with the knowledge approach of pro/con lists~\cite{kriplean2012supporting}. The temporal reflector follows a rather novel reflection approach by emphasizing the need to organize one's past personal experience~\cite{phemister2017leibniz}. We noticed that participants readily shared their personal accounts related to the issue, suggesting that temporal prompts effectively encouraged users to recount their experiences.

In this way, these reflectors serve as catalysts for introspection, motivating users to contemplate their thoughts, perspectives, and experiences. They encourage users to ponder relevant questions and seek answers pertaining to the discussion topic. Importantly, they expand users' considerations on the issue, encouraging them to think beyond their personal biases and viewpoints. Consequently, the adoption of reflectors can enhance deliberativeness as users are incentivized to invest time and effort in exploring the nuances of the discussion topic. 

Future work could also explore how the reflectors dynamically respond to users' traits (i.e., TK-TI and SRIS) to enhance the personalization of reflection. For example, users with higher TK-TI might benefit from reflectors tailoring prompts to more advanced or nuanced aspects of the topic. For those with lower TK-TI, the textual prompts could be adapted to provide foundational information. Likewise, users with a higher SRIS may respond well to prompts that encourage deeper introspection, whereas those with a lower SRIS might require prompts with clearer guidance or cues to stimulate reflective thinking.

\subsubsection{Reflectors promote willingness to express opinions} 
\label{sec: reflectors promote opinion expression}
Consistent with the work by Zhang et al.~\cite{zhang2021nudge}, our findings revealed that the reflectors promote willingness to express opinions. In addition, our study revealed that the persona reflector, in particular, encouraged users to be more opinionated and willing to express their viewpoints even in a public forum. This is noteworthy as public opinion expression directly contributes to political discourse\cite{zhang2021nudge} and constitutes an indispensable element of the deliberation process~\cite{macedo1999deliberative, zhang2014perceived}. As engaging in public discourse has higher barriers such as a potentially hostile environment~\cite{zhang2021nudge} and is dependent on affective variables like fear of isolation~\cite{sullivan1999psychological, zhang2021nudge}, the use of reflectors like persona can serve as a catalyst for public opinion expression, especially among individuals who are hesitant to share such experiences in a public forum~\cite{sullivan1999psychological}. A similar effect can be expected from temporal prompts as they encourage users to share their personal experiences related to the issue.  

\subsubsection{Reflectors are impartial guides} Our findings from study 1 revealed that participants perceived reflectors as neutral, unbiased, and objective guides for their reflective thinking, distinguishing them from the influence of other users' comments (see section~\ref{sec: users comments vs reflectors}). This suggests that users may engage in more objective and balanced deliberation when supported by reflectors, ultimately leading to a more thorough and impartial discussion of the issue.

Overall, our findings highlight the instrumental role of reflectors in facilitating users' self-reflection and improving deliberative quality. Leveraging on the use of reflectors to elevate the value and impact of individual contributions in discussions would enrich the quality and depth of discourse on online deliberative platforms.

\subsection{Different Reflectors Suit Different Online Deliberation Environments}
\subsubsection{Persona: For deliberative environments that seek to promote balanced and opinionated viewpoints} Participants in the persona condition demonstrated a tendency to provide viewpoints that were both more balanced and opinionated. This aligns with what is often referred to as the \textit{persona effect}~\cite{lester1997persona} where the presence of a character, representing diverse demographics and backgrounds, \textbf{appear to psychologically distance users from their own beliefs}~\cite{lester1997persona}, thus encouraging them to explore a wider range of opinions. Moreover, users in this condition seemed to feel less social pressure to conform to dominant opinions, thus becoming more willing to express their own personal views (see section~\ref{sec: reflectors promote opinion expression}. This, in turn, led to discussions that were both more open and opinionated.

\subsubsection{Storytelling: For deliberative environments that seek to solicit feedback on complex and sensitive topics} In the case of storytelling, the results showed a somewhat middle-ground approach. Participants from study 1 noted that storytelling could be viewed as an extended version of persona, albeit delivering perspectives in a more indirect manner, by including additional contextual information and background details. While some participants appreciated this, others found it \textbf{time-consuming to read} and preferred the succinctness of the persona reflector. Despite so, storytelling can be particularly valuable when soliciting feedback and opinions on complex or sensitive issues and on topics that are unfamiliar to the users. This is because stories excel at providing valuable contextual information~\cite{mcdowell2021storytelling}, helping users to better understand intricate issues~\cite{andrews2009storytelling, billings2016storytelling} and making abstract or sensitive subjects more relatable~\cite{billings2016storytelling}.

\subsubsection{Temporal Prompts: For deliberative environments that seek to promote individualized viewpoints} As for temporal prompts, participants exhibited greater inclination to share their personal experiences. This could be advantageous in scenarios where the deliberative environment seeks to elicit more individualized viewpoints or personal accounts related to a particular issue. It is important to note, though, that \textbf{opinions expressed under temporal prompts may lean somewhat towards being one-sided}. Nevertheless, temporal prompts can stimulate participants' contemplation on the topic through the lens of their own experiences. This tendency likely accounts for the observed increase in participants' thinking and the amplification of their knowledge base as shown in our qualitative results. This, in turn, might explain why the temporal prompts reflector exhibits the highest argument diversity.

\paragraph{\textnormal{Overall, our study underscores that the choice of reflectors can significantly influence the dynamics of online discussions through shaping users' own expressions. Each reflector presents distinct advantages and can be strategically employed to shape the nature and quality of online deliberations according to the specific desired outcomes for online platforms or discussions.}} 

\subsection{Internal vs. External Self-Reflection Approaches - The Case of Temporal Prompts and Persona}
Temporal prompts encouraged users to reflect on their personal experiences, memories and the passage of time. This form of \textbf{internal} reflective mechanism helps individuals to gain insights into their own beliefs, values, and personal experiences~\cite{morin1990inner} and to connect their current opinions or perspectives with their past, facilitating a deeper understanding of how their thoughts have evolved over time. This self-awareness~\cite{morin1990inner} is fundamental for constructing more thoughtful and nuanced arguments and opinions.

Persona on the other hand, encouraged users to adopt different viewpoints, by assuming the role of someone with distinct characteristics, beliefs, or experiences~\cite{blomkvist2002persona}. This form of \textbf{external} reflective mechanism challenges individuals to think beyond their personal biases and preconceived notions~\cite{arbinger2016outward} and to empathize with others and appreciate the complexity of issues from different angles. This type of reflective mechanism is essential for fostering open-mindedness, promoting constructive dialogue, and enhancing deliberativeness.

Both mechanisms facilitated by temporal prompts and persona reflectors may serve \textbf{complementary roles} in the reflective process. For example, temporal prompts can serve as an initial step in the reflective process to encourage users to reflect on their own experiences and beliefs, providing a solid foundation of self-awareness. Persona can then be introduced to expand their horizons to help individuals to explore alternative viewpoints, fostering open-mindedness and a willingness to consider diverse perspectives. Together, they can create a \textbf{balanced approach to self-reflection}, to allow users to better equipped to engage in deliberation. 

In future studies, we may want to look into the combinations and pairing of different reflectors to yield optimal outcomes that are tailored to the needs of specific deliberative environments. Exploring the synergy between different reflectors holds promise in mitigating any potential drawbacks associated with a single reflector approach and such hybrid approaches could potentially further enhance the overall quality and depth of online discussions.

\subsection{Potential of Utilizing LLM to Support Self-Reflection}
Unlike conventional online deliberation platforms where users formulate their opinions through a combination of user-driven self-reflection and the consumption of other users' comments on the same issue, our study has illustrated how leveraging LLM can aid this process through unobtrusive interface nudges. This highlights the potential for LLMs not just as tools for assisting in writing but also as facilitators of reflective thinking, guiding users toward decisions grounded in thorough consideration rather than hasty, uninformed judgments.
Nevertheless, it's crucial to acknowledge that the effectiveness of LLMs in supporting deliberation hinges on the quality of the responses they provide. While we employed GPT-3.5 in this work, we believe that future iterations and new LLMs (e.g., GPT-4) should yield improved responses.

\subsection{Using Reflectors in Non-Deliberative Contexts}
While the task presented in this paper was specific and done in an online deliberation context, the effects of the reflectors could be leveraged in other contexts. 

\paragraph{Online Forums.}
The persona reflector can be suitable for online discussion platforms (e.g., Reddit and Quora) as we found that it enhances the overall deliberative quality. Being a subtle nudge, online discussion platforms can easily leverage its use as a plugged in tool to facilitate users' reflection and opinion formulation prior to posting. Additionally, depending on the discussion topic, the use of the storytelling reflector could be employed on sensitive or controversial issues, augmenting users understanding with contextual information. 

\paragraph{Online Review Systems.}
The use of the temporal prompts reflector could be implemented on review interfaces (e.g., Google reviews, TripAdvisor, Yelp and Rotten Tomatoes) by nudging users with reflective questions while they are crafting their reviews. As we found that temporal prompts lead to more extensive sharing of personal experiences, such reflective nudges could enhance the quality and authenticity of reviews. This in turn benefits readers by providing them with richer and more informative content for making informed decisions.

\paragraph{Beyond User-Generated Content Platforms to Other Domains}
The reflectors designed for online deliberation can be adapted in educational contexts to guide students in navigating complex topics. For instance, the storytelling reflector could facilitate discussions on intricate subjects, while the temporal prompts reflector may encourage students to reflect on their learning experiences, enhancing their critical thinking.

Moreover, the reflectors can be tailored for use in personal exploration within counseling or mental health applications, providing a supportive tool for individuals to delve into their emotions and thoughts. Future research endeavors could explore the cross-domain generalizability of these findings, seeking to understand the reflectors' boundaries and flexibility in diverse domains. 
\section{Limitations}
There are limitations in this work. Firstly, study 1 was conducted on young adults. Although this is a common target population for online deliberation platforms, these platforms are widely used in a range of populations. We mitigated this issue in study 2 with a larger and more diverse participants pool. Future work may focus on understanding the cultural aspects of self-reflection when using the different reflectors from different populations. Assessing the reflectors and evaluating their merits for different populations is an important goal for future use and research on self-reflection on online deliberation platforms. 

The task employed in both studies was specific, aiming to achieve depth and clarity in understanding the influences of diverse reflection approaches within a well-defined context~\cite{slack2001establishing}. While this establishes internal validity, it may conflict with the broader goal of generalizability and ecological validity. It is important to note that assessing ecological validity relies on first establishing internal validity~\cite{cahit2015internal, slack2001establishing, campbell2015experimental, cook1979quasi}. Therefore, having a focused investigation on a specific issue allows us to have a detailed examination of the reflection nudges on deliberativeness. Future work could extend the application of the reflective nudges to explore other contentious topics with varying complexity and nature.
\section{Conclusion}
In this work, we investigated how various interface-based reflection nudges impacts the quality of deliberation. We provide the missing link between the two by examining five text-based reflective nudges: persona, analogies and metaphors, cultural prompts, storytelling and temporal prompts, across two studies. In study 1, we identified persona, storytelling and temporal prompts as the preferred nudges for implementation on online deliberation platforms. In study 2, we discussed how different reflective nudges can significantly shape the dynamics of online discussions. Our results expand current work on self-reflection and online discussions, offering insights into how different self-reflection approaches support the deliberation process. Thus, providing valuable guidance on the use of reflectors on online deliberation platforms.

\bibliographystyle{ACM-Reference-Format}
\bibliography{reference.bib}

\newpage
\appendix

\label{appendix: reflectors selection}

\begin{table*}[!htbp]
\caption{Detailed Descriptions, Rationales and Applications of the Selected Reflectors in Study 1}
\label{tab: reflectors selection}
\centering
\scalebox{0.60}{
\begin{tblr}{
  width = \linewidth,
  colspec = {Q[200]Q[750]Q[900]},
  row{1} = {c},
  cell{2}{1} = {c},
  cell{3}{1} = {c},
  cell{4}{1} = {c},
  cell{5}{1} = {c},
  cell{6}{1} = {c},
  hlines,
  vlines,
}
\textbf{Reflector} & \textbf{Description and Rationale} & \textbf{Applications} \\
\textbf{Persona} 
& {\labelitemi\hspace{\dimexpr\labelsep+0.5\tabcolsep}The concept of persona in the HCI community was originally introduced by Cooper ~\cite{cooper1999inmates, cooper2007face}.\\\labelitemi\hspace{\dimexpr\labelsep+0.5\tabcolsep}A persona is a fictional character that helps individuals to empathize by understanding the needs, personal goals and personal contexts ~\cite{chang2008personas, blomkvist2002persona}.\\\labelitemi\hspace{\dimexpr\labelsep+0.5\tabcolsep}In adopting the persona's perspective, individuals engage in a form of perspective-taking, allowing them to think in the shoes of an imagined other's viewpoint and experiences. \\\labelitemi\hspace{\dimexpr\labelsep+0.5\tabcolsep}The use of personas is also found to be beneficial in understanding and designing for different stakeholders ~\cite{chang2008personas}.} & {\labelitemi\hspace{\dimexpr\labelsep+0.5\tabcolsep}Traditionally, persona is being used by practicing designers in interaction design ~\cite{chang2008personas}.\\\labelitemi\hspace{\dimexpr\labelsep+0.5\tabcolsep}In the work of PolicyScape ~\cite{kim2019crowdsourcing}, the authors employ the perspective-taking approach to help users to consider different stakeholders’ perspectives on policy issues.\\\labelitemi\hspace{\dimexpr\labelsep+0.5\tabcolsep}Other applications ~\cite{zhang2021nudge} include considering viewpoints of an imagined other or writing a narrative of an imagined perspective.\\\labelitemi\hspace{\dimexpr\labelsep+0.5\tabcolsep}It is also employed in areas that examine interactions with other people, to enable individuals to learn about personal differences ~\cite{vink2022building}.} \\
{\textbf{Analogy and Metaphor}\\\textbf{ }} & {\labelitemi\hspace{\dimexpr\labelsep+0.5\tabcolsep}Analogy and metaphor are figures of speech that are used to suggest a resemblance between two disparate concepts ~\cite{Webster,lee2007causal,gentner1982scientific}.\\\labelitemi\hspace{\dimexpr\labelsep+0.5\tabcolsep}Their purpose is to convey a deeper meaning, create an imaginative image or for clarifying a concept ~\cite{gentner1982scientific}.\\\labelitemi\hspace{\dimexpr\labelsep+0.5\tabcolsep}Analogy and metaphor are prominent in theoretical discussions to allow one to comprehend and reflect on their own inner thoughts ~\cite{vink2022building}, relating to one’s internal reflection ~\cite{creed2020institutional}.\\\labelitemi\hspace{\dimexpr\labelsep+0.5\tabcolsep}They have shown to help individuals re-frame their thoughts ~\cite{vink2022building} but in some cases can become too abstract or complex ~\cite{vink2022building,schwind2009metaphor}.}  & {\labelitemi\hspace{\dimexpr\labelsep+0.5\tabcolsep}Vink and Koskela-Huotari ~\cite{vink2022building} employ analogy and metaphor to support service designers when engaging with caregivers in their cognitive thinking and reflection.\\\labelitemi\hspace{\dimexpr\labelsep+0.5\tabcolsep}Keefer and Landau ~\cite{keefer2016metaphor} employ analogy and metaphor to improve reasoning and to solve real world problems.\\\labelitemi\hspace{\dimexpr\labelsep+0.5\tabcolsep}Schwind ~\cite{schwind2009metaphor} uses metaphor-reflection in the healthcare context to explore how such narrative reflection assist patients in exploring the meaning of their illness events.\\\labelitemi\hspace{\dimexpr\labelsep+0.5\tabcolsep}Niebert et al. ~\cite{niebert2012understanding} employ imaginative thinking with the use of metaphors and analogies in everyday life to teach science.} \\
\textbf{Cultural Prompt} & {\labelitemi\hspace{\dimexpr\labelsep+0.5\tabcolsep}Reflective questions to cultivate cultural awareness ~\cite{adams2003reflexive}.\\\labelitemi\hspace{\dimexpr\labelsep+0.5\tabcolsep}Involves considering one’s own perspectives, beliefs, biases and experiences in relation to the environment they are in ~\cite{adams2003reflexive}.\\\labelitemi\hspace{\dimexpr\labelsep+0.5\tabcolsep}This process of self-awareness has shown to allow individuals to understand the factors influencing their perceptions and interactions ~\cite{vink2022building, adams2003reflexive}.} & {\labelitemi\hspace{\dimexpr\labelsep+0.5\tabcolsep}Applications on cultivating cultural awareness have largely been focused on teachers and learners. \\\labelitemi\hspace{\dimexpr\labelsep+0.5\tabcolsep}Civitillo et al. ~\cite{civitillo2019interplay} found that teachers who were more culturally responsive also showed more self-reflection on their own teaching.\\\labelitemi\hspace{\dimexpr\labelsep+0.5\tabcolsep}Donati ~\cite{donati2011modernization} and Mouzelis ~\cite{mouzelis2010self} showed that cultural awareness involves reflecting on the contexts of family, state, market and other cultural contexts.} \\
\textbf{Storytelling} & {\labelitemi\hspace{\dimexpr\labelsep+0.5\tabcolsep}Storytelling involves conveying thoughts, convictions, personal encounters, and life teachings using vivid narratives that stir deep emotions and provoke profound insights~\cite{serrat2008storytelling}.\\\labelitemi\hspace{\dimexpr\labelsep+0.5\tabcolsep}Composing stories is a promising strategy for achieving high quality reflection ~\cite{hamilton2019digital}.\\\labelitemi\hspace{\dimexpr\labelsep+0.5\tabcolsep}Multiple papers have looked at the relationship between storytelling, learning and reflection.} & {\labelitemi\hspace{\dimexpr\labelsep+0.5\tabcolsep}Work centred on storytelling have developed rubrics for assessing how students used digital storytelling format to reflect on their study abroad experiences ~\cite{hamilton2019digital}.\\\labelitemi\hspace{\dimexpr\labelsep+0.5\tabcolsep}Freidus and Hlubinka ~\cite{freidus2002digital} present the use of digital storytelling in community development settings to promote reflective practice and found that individuals articulate better in their thought process.} \\
\textbf{Temporal Prompt} & \labelitemi\hspace{\dimexpr\labelsep+0.5\tabcolsep}Reflective questions that involves the aspect of time ~\cite{phemister2017leibniz, vink2022building, dawson2014reflections}. It involves prompting one’s life experiences, examining one’s history and past events and how they have evolved over time ~\cite{phemister2017leibniz, vink2022building, dawson2014reflections}. & 
{\labelitemi\hspace{\dimexpr\labelsep+0.5\tabcolsep}Smallwood et al. ~\cite{smallwood2011self} investigated how self-reflection impacts both retrospection (reflecting on the past) and prospection (reflecting on the future).\\\labelitemi\hspace{\dimexpr\labelsep+0.5\tabcolsep}Lindström et al. ~\cite{lindstrom2006affective} designed affective diary to allow users to reflect on their inner thoughts while recording experiences of past events and piecing their own stories.\\\labelitemi\hspace{\dimexpr\labelsep+0.5\tabcolsep}Kyoko and Jacobs~\cite{murakami2017connecting} observed how self-reflecting on the past can build a shared future and knowledge in the context of family reminiscence.}
\end{tblr}
}
\Description{This table provides comprehensive information on the five selected reflectors used in Study 1. The table comprises of three columns with headers: reflector, description and rationale and applications. Under description and rationale, each reflector is explained in terms of its origin, definition, and underlying rationale. The applications column outlines the contexts and domains where each reflector is employed to facilitate reflective thinking, supported by relevant references.}
\end{table*}

\newpage
\label{appendix: prompt engineering}

\begin{table*}[!htbp]
\caption{The prompts utilized in \textit{Help Me Reflect} along with the corresponding constraints for each reflector when communicating with ChatGPT API to produce textual reflective nudges. According to ChatGPT API specification, ``system role'' specifies the role the model assumes in the session, and ``user input'' provides the model with the user's input prompt.}
\label{tab: prompt engineering}
\scalebox{0.77}{
{\begin{tabular}{@{}ll@{}}
\toprule
\textbf{Reflective Nudges} & \textbf{Prompt Template} \\ 
\midrule
\textit{For all reflective nudges} & \begin{tabular}[c]{@{}l@{}}$\bullet$ \textbf{system role}: You are a helpful assistant focusing on supporting users' self-reflection on a given topic;\\ $\bullet$ \textbf{user input}: Topic: [topic]. \textit{(The following continues the user input for each of the reflective nudges.)}\end{tabular} \\
\hline 
Personas & \begin{tabular}[c]{@{}l@{}} For the above topic, create six distinct personas representing different perspectives \\ on the topic. Provide the name, age and occupation for each persona. \\ [1ex] Here is the format of the results: \\ 1. [Name1], [Age1], [Occupation1], [Perspective1] \\ 2.  [Name2], [Age2], [Occupation2], [Perspective2]\\ ... \\ [1ex] Requirements: \\ Create three male personas and three female personas; \\ For each perspective, be concise, giving at most two sentences; \\ No duplicates; \\ Six distinct versions only \end{tabular} \\
\hline
Storytelling & \begin{tabular}[c]{@{}l@{}} Generate six distinct stories on the topic matter. \\ [1ex] Here is the format of the results: \\ Story1: \\ Story2: \\ ... \\ [1ex] Requirements: \\ Create three stories with a positive tone and three stories with a negative tone; \\ Create three male protagonists and three female protagonists; \\ No duplicates or similar story line; \\ Six distinct versions only \\ [1ex] * Depending on the length of the story generated, we prompt GPT to either \\ lengthen (\textit{Extend Story1}) or condense (\textit{Make Story1 concise}) the narrative. \end{tabular} \\ 
\hline
Analogies and Metaphors & \begin{tabular}[c]{@{}l@{}} Generate six distinct analogies and metaphors on the topic matter. \\ [1ex] Here is the format of the results: \\ Version1: \\ Version2: \\ ... \\ [1ex] Requirements: \\ No duplicates or similar ones; \\ Six distinct versions only \end{tabular} \\ 
\hline
Temporal Prompts & \begin{tabular}[c]{@{}l@{}} Generate six distinct reflective questions on one's life journey and experiences on the \\ topic matter. \\ [1ex] Here is the format of the results: \\ Version1: \\ Version2: \\ ... \\ [1ex] Requirements: \\ No duplicates or similar ones; \\ Six distinct versions only \end{tabular} \\ 
\hline
Cultural Prompts & \begin{tabular}[c]{@{}l@{}} Generate six distinct reflective questions on one's culture, customs, biasness and \\ assumptions on the topic matter. \\ [1ex] Here is the format of the results: \\ Version1: \\ Version2: \\ ... \\ [1ex] Requirements: \\ No duplicates or similar ones; \\ Six distinct versions only \end{tabular} \\
\bottomrule
\end{tabular}}
}
\Description{This table details the prompts used to generate the textual reflective nudges, along with the corresponding constraints applied when communicating with the ChatGPT API. The table has two columns with headers: reflective nudges and prompt template. The prompt template column provides a breakdown of the specific prompts utilized for creating each of the five textual reflective nudges. These prompts adhere to the template established by White et al., which entails defining a task, incorporating constraints, and setting clear expectations for the generated output.}
\end{table*}

\begin{table*}[!htbp]
\caption{Rationales of the constraints set out for GPT}
\label{tab: prompt constraint rationale}
\centering
\scalebox{0.74}{
\begin{tblr}{
  width = \linewidth,
  colspec = {Q[352]Q[588]},
  row{1} = {c},
  hlines,}
\textbf{Constraints} & \textbf{Rationale} \\
{(Persona) Create three male personas and three female personas. \\
(Storytelling) Create three male protagonists and three female protagonists. \\~} & {\labelitemi\hspace{\dimexpr\labelsep+0.5\tabcolsep}Scholars have found that large language models such as GPT-3 produce gender stereotypes and biases~\cite{brown2020language,lucy2021gender, huang2019reducing, nozza2021honest, johnson2022ghost} \\\labelitemi\hspace{\dimexpr\labelsep+0.5\tabcolsep} Hence, this constraint was set out to mitigate any potential gender imbalances as the discussion topic is a contentious one. \\\labelitemi\hspace{\dimexpr\labelsep+0.5\tabcolsep} Ensures GPT does not provide viewpoints that gives preferential treatment of one gender over the other.} \\
(Persona) For each perspective, be concise, giving at most two sentences. & {\labelitemi\hspace{\dimexpr\labelsep+0.5\tabcolsep}This constraint was introduced after observing GPT's tendency to generate long perspectives during testing. \\\labelitemi\hspace{\dimexpr\labelsep+0.5\tabcolsep} Ensures GPT deliver concise perspectives.} \\
(Storytelling) Of the six stories, create three with a positive tone and three with a negative tone. & \labelitemi\hspace{\dimexpr\labelsep+0.5\tabcolsep} Ensures that the reflector captures a broad spectrum of impacts, including both positive and negative aspects to avert one-sidedness. \\
(Storytelling) Extend or make concise. & {\labelitemi\hspace{\dimexpr\labelsep+0.5\tabcolsep} Ensures a well-rounded representation of stories two long stories, two short stories, and two stories of medium length. \\\labelitemi\hspace{\dimexpr\labelsep+0.5\tabcolsep} This strategy allows us to capture the full spectrum of narrative possibilities and thereby provides a more thorough analysis of the reflector's effectiveness.} \\
Here is the format of the results & \labelitemi\hspace{\dimexpr\labelsep+0.5\tabcolsep} Ensures that GPT provides results in a consistent format.      
\end{tblr}}
\Description{This table provides a breakdown of the constraints applied to GPT during the generation of textual reflective nudges, along with the underlying rationales for each constraint. The table has two columns with column header: constraints and rationale. In the constraints column, we list the specific constraints implemented for each reflector. Meanwhile, the rationale column elucidates the motivations behind the establishment of these constraints.}
\end{table*}

\label{appendix: example output persona}

\begin{figure*}[!htbp]
  \centering
  \includegraphics[scale=0.42]{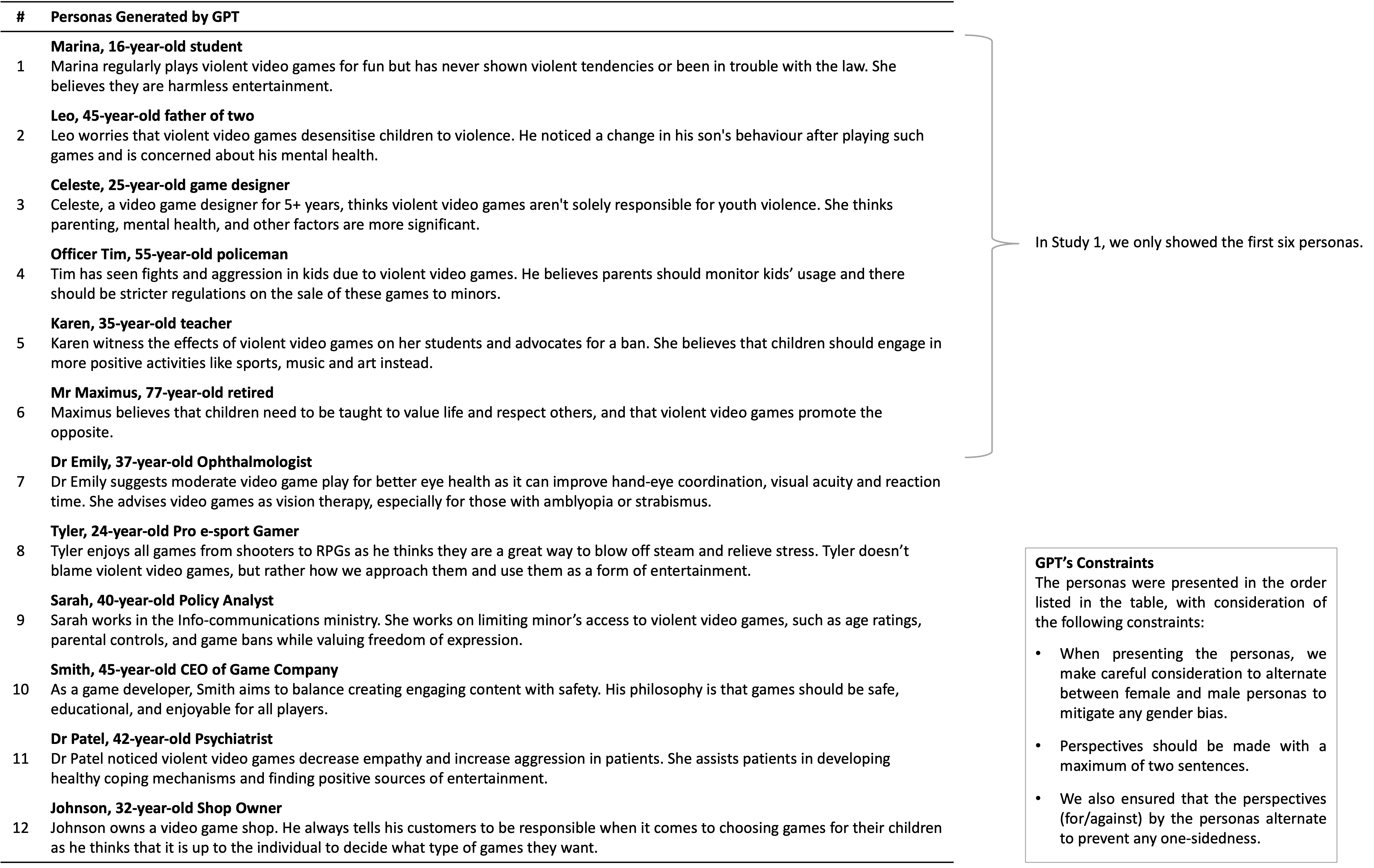}
  \caption{Example output of the reflective nudge - persona. Full list of personas generated by GPT with illustration to the constraints set out for GPT.}
  \label{fig: GPT persona output}
  \Description{A comprehensive list of the personas generated by GPT with the first six personas in Study 1. The personas were presented in consideration with the following three constraints: alternating males' and females' perspectives to mitigate any potential gender imbalances, concise perspectives having at most two sentences and alternating perspectives of for and against to prevent any one-sidedness.}
\end{figure*}

\newpage

\begin{figure*}[!htbp]
  \centering
  \includegraphics[width=0.4\linewidth]{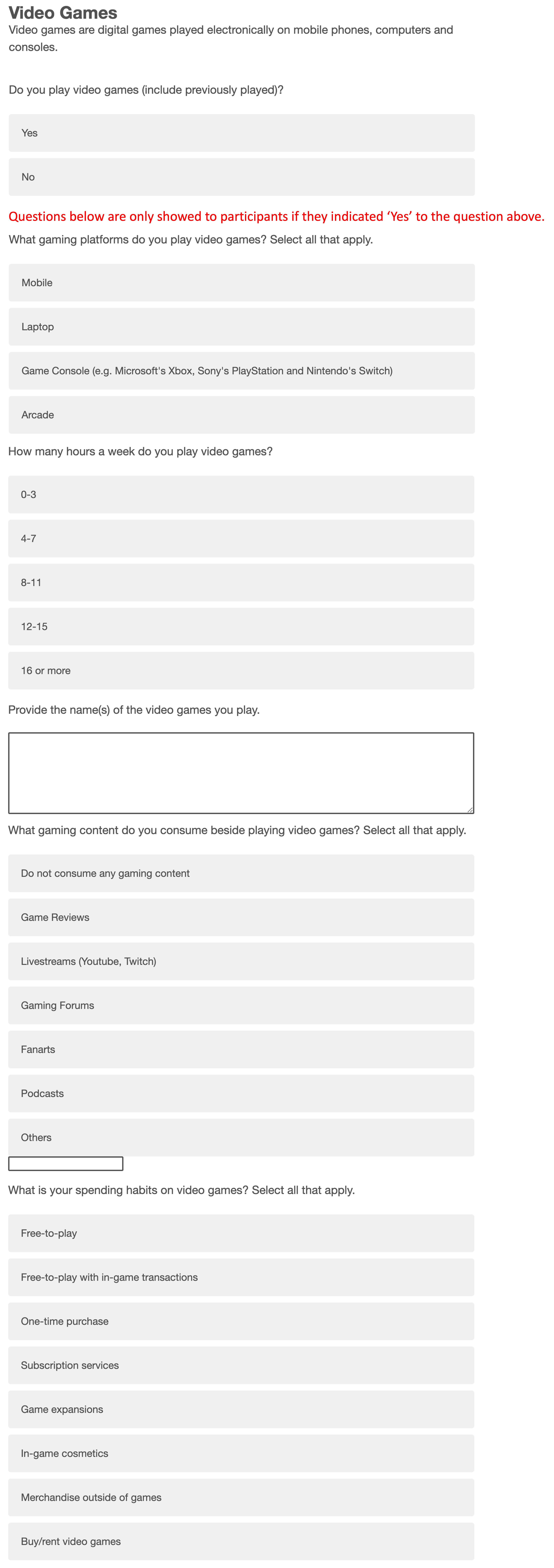}
  \caption{Complete list of TK-TI Questionnaire, covering questions assessing participants' video game knowledge (e.g., if they have gaming experience and if they could name some examples of video games) and their interest in the topic (e.g, habitual attitudes and caring on gaming content preferences and spending habits).}
  \label{fig: TK-TI Questionnaire}
  \Description{Complete list of TK-TI Questionnaire, covering six questions on participants' video game knowledge and their interest in the topic. They include: 1: Do you play video games (include previously played)? Options include `yes' and `no'. The following questions were only shown if participants indicated `yes' to the previous question: 2: What gaming platforms do you play video games? Select all that apply. This question include the following options: mobile, laptop, game console (e.g., Microsoft's Xbox, Sony's PlayStation and Nintendo's Switch) and Arcade; 3: How many hours a week do you play video games? Options include 0-3, 4-7, 8-11, 12-15 and 16 or more; 4: Provide the name(s) of the video games you play. 5: What gaming content do you consume beside playing video games? Select all that apply. Options include do not consume any gaming content, games reviews, livestreams (YouTube, twitch), gaming forums, fanarts, podcasts and others; 6: What is your spending habits on video games? Select all that apply. Options include free-to-play, free-to-play with in-game transactions, one-time purchase, game expansions, in-game cosmetics, merchandise outside of games and buy/rent video games. Questions 1 and 4 assess participants' knowledge while questions 2, 3, 5 and 6 inquired about their interest, specifically on their habitual attitudes and caring on the topic.}
\end{figure*}

\newpage
\label{sec: st1-demo}

\begin{table*}[!htbp]
\caption{Demographic Profiles of Participants in Study 1. Note: Although not included in the primary analysis, we collated participants' topic knowledge and topic interest (TK-TI) scores, as well as their self-reported self-reflection and insight scale (SRIS) scores, to account for any potential fixed effects associated with these two covariates. However, it's worth noting that both covariates showed no statistically significant impact on the rankings of the reflectors. Furthermore, we found no meaningful correlations between participants' TK-TI and SRIS scores and their interactions with any of the reflectors.}
\label{tab: st1-demo}
\begin{tabular}{@{}cccccc@{}}
\toprule
Participants & Gender & Age & Education & TK-TI Scores & SRIS scores \\ 
\midrule
P1 & Male & 28 & Bachelor's Degree & 5 & 97 \\
P2 & Female & 21 & Undergraduate & 6 & 77 \\
P3 & Male & 20 & Undergraduate & 6 & 94 \\
P4 & Female & 20 & Undergraduate & 0 & 119 \\
P5 & Female & 20 & Undergraduate & 4 & 82 \\
P6 & Male & 36 & Post-graduate Degree & 5 & 78 \\
P7 & Female & 20 & Undergraduate & 6 & 86 \\
P8 & Female & 24 & Bachelor's Degree & 0 & 104 \\
P9 & Male & 19 & Undergraduate & 0 & 109 \\
P10 & Female & 22 & Bachelor's Degree & 0 & 90 \\
P11 & Female & 26 & Post-graduate Degree & 5 & 87 \\
P12 & Female & 23 & Bachelor's Degree & 6 & 75 \\
\bottomrule
\end{tabular}
\Description{This table provides a comprehensive overview of the demographic characteristics of the participants in Study 1. The table is organized into six columns with the following column headers: Participants, Gender, Age, Education, TK-TI Scores, and SRIS Scores. The participants column enumerates each participant, while the gender column indicates whether they are male or female. In the education column, the participants' educational backgrounds are categorized as either having a bachelor's degree, undergraduate education, or post-graduate degree.}
\end{table*}

\newpage
\label{appendix: study 2 demographics}

\begin{table*}[!htbp]
\caption{Summary Statistics of the Demographics Profile in Study 2. Note: Although not included in the primary analysis, we collated participants' demographic profile, topic knowledge and topic interest (TK-TI) scores, as well as their self-reported self-reflection and insight scale (SRIS) scores, to account for any potential fixed effects associated with these three covariates. It is worth noting that all covariates showed no statistically significant impact on the measurements of deliberative quality with the exception of argument repertoire (see section~\ref{sec: quantitative}). Furthermore, we found no meaningful correlations between participants' TK-TI and SRIS scores and their interactions with any of the reflectors.}
\label{tab: demograhics study 2}
\begin{tabular}{@{}llcccc@{}}
\toprule
\multicolumn{2}{c}{\textbf{Demographics Profile}} & \textbf{Control} & \textbf{Persona} & \textbf{Storytelling} & \textbf{Temporal Prompts} \\ 
\midrule
\textbf{Participants} & \begin{tabular}[c]{@{}l@{}} Total number of participants \\ (after excluding outliers) \end{tabular} & 29 & 30 & 28 & 30 \\
\hline
\multirow{3}{*}{\textbf{Gender}} & Number of Males & 21 & 18 & 18 & 15 \\
& Number of Females & 7 & 11 & 10 & 14 \\
& Prefer Not to Say & 1 & 1 & 0 & 1 \\
\hline
\textbf{Age} & Mean Age & 38.9 & 37.1 & 34.6 & 44.3 \\
\hline
\multirow{6}{*}{\textbf{Ethnicity}} & Asian or Pacific Islander & 5 & 6 & 7 & 7 \\
& Black or African American & 0 & 3 & 0 & 1 \\
& Hispanic or Latino & 3 & 3 & 1 & 3 \\
& Multiracial or Biracial & 0 & 0 & 0 & 0 \\
& Native American or Alaska Native & 1 & 0 & 3 & 2 \\
& White or Caucasian & 20 & 18 & 17 & 17 \\
\hline
\multirow{4}{*}{\textbf{Education}} & \begin{tabular}[c]{@{}l@{}} High School, Diploma \\ or the equivalent \end{tabular} & 4 & 8 & 3 & 3 \\
& Associate's Degree & 0 & 3 & 2 & 4 \\
& Bachelor's Degree & 16 & 15 & 18 & 21 \\
& Post-graduate Degree & 9 & 4 & 5 & 2 \\
\hline
\textbf{Time} & Mean Completion Time (mins) & 11.5 & 24.0 & 37.1 & 36.8 \\
\hline
\textbf{TK-TI} & TK-TI Scores ($M \pm S.D.$) & $5.07 \pm 1.85$ & $5.57 \pm 1.17$ & $5.68 \pm 0.55$ & $5.40 \pm 1.22$\\
\hline
\textbf{SRIS} & SRIS Scores ($M \pm S.D.$) & $84.3 \pm 15.1$ & $86.6 \pm 20.2$ & $87.1 \pm 15.3$ & $91.2 \pm 15.2$ \\
\bottomrule
\end{tabular}
\Description{This table provides a comprehensive summary of the demographic profiles of the participants in Study 2. We gathered information about participants genders, age, ethnicity, education. Overall the four groups are homogeneous.}
\end{table*}

\newpage
\label{appendix: post-task feedback}

\begin{figure*}[!htbp]
  \centering
  \includegraphics[width=\linewidth]{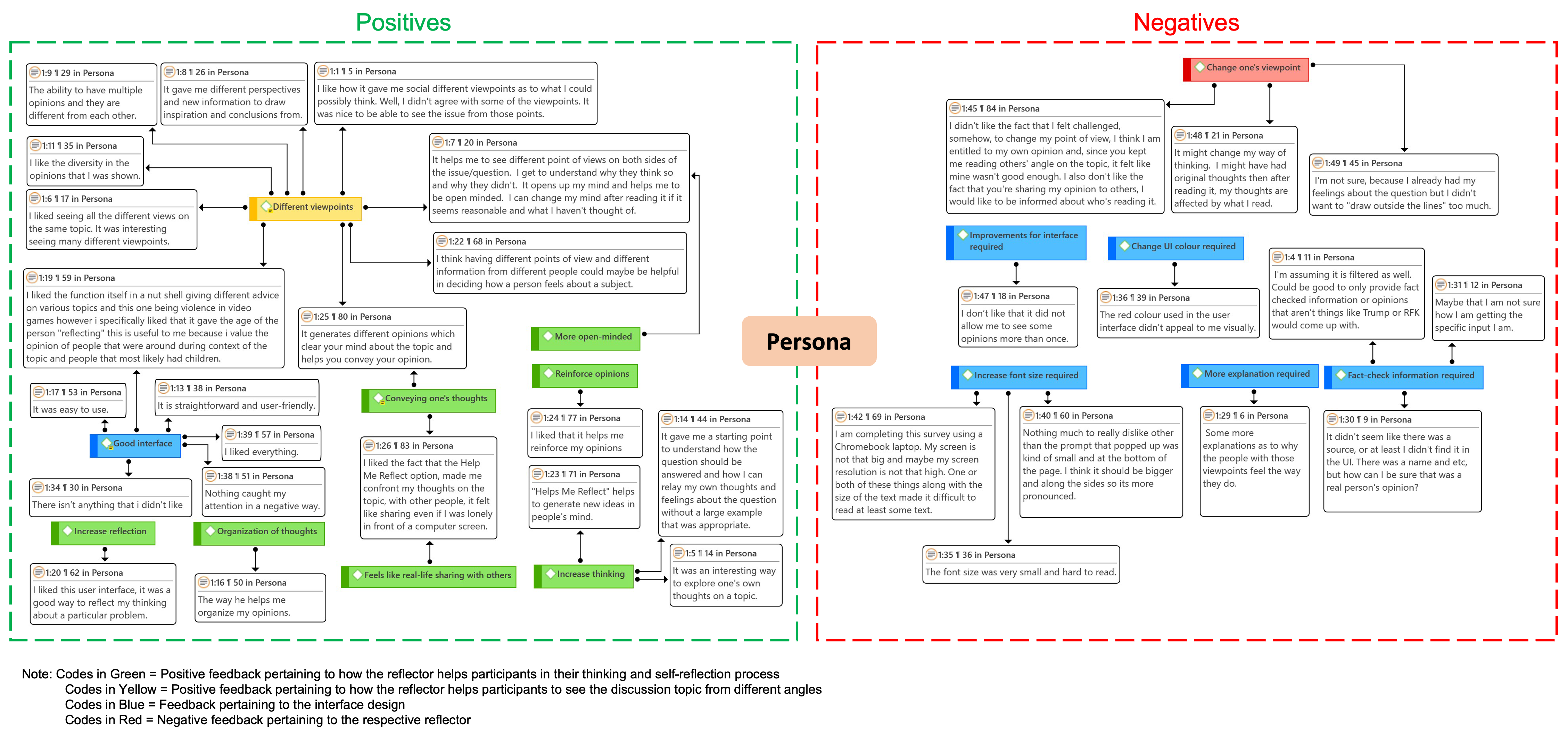}
  \caption{Participants' post-task feedback for the reflector - Persona. The figure provides a snapshot of the positive and negative aspects the reflector brought to the participants' experiences during their internal reflection process. Throughout the coding process, multiple codes could be applied to each participant's feedback.}
  \label{fig: persona post-task}
  \Description{Participants' Feedback on the Persona Reflector. We present an overview of participants' feedback regarding the Persona Reflector. Feedback are categorized into two main sections: Positives and Negatives, each represented with distinctive color codes for clarity. Green: Positive feedback highlighting how the Persona Reflector assists participants in their thinking and self-reflection process. Yellow: Positive feedback indicating how the Persona Reflector helps participants explore the discussion topic from various perspectives. Blue: Feedback related to the interface design of the Persona Reflector. Red: Negative feedback pointing out issues or challenges participants encountered while using the Persona Reflector.}
\end{figure*}

\begin{table*}[!htbp]
\caption{Full List of the Post-Task Feedback for Persona: The numbers represent the frequency of occurrence of specific feedback mentioned by participants. The feedback are listed in descending order, from the highest to the lowest.} 
\label{tab: pesona post-task feedback}
\begin{tabular}{@{}cccc@{}}
\toprule
\multicolumn{2}{c}{\textbf{Positives}} & \multicolumn{2}{c}{\textbf{Negatives}} \\ 
\midrule
Different viewpoints & 16 & Fact-check information & 4 \\
Good interface & 10 & More explanation & 3 \\
Heightened thinking & 4 & Increase font size & 3 \\
Conveying one's thoughts & 2 & Change one's viewpoint & 3 \\
Organization of thoughts & 2 & Change UI colour & 1 \\
Reinforce opinion & 1 & Improvements for interface & 1 \\
Increase reflection & 1 &  & \\
Feels like real-like sharing with others & 1 &  & \\
More open-minded & 1 &  & \\
\bottomrule
\end{tabular}
\Description{This table is an extension of the top three post-task feedback for the persona reflector. We present the full List of the post-task feedback, both positives and negatives feedback for persona. We also report the numbers representing the frequency of occurrence of specific feedback mentioned by participants. The feedback are listed in descending order, from the highest to the lowest occurrence.}
\end{table*}

\newpage
\begin{figure*}[!htbp]
  \centering
  \includegraphics[width=\linewidth]{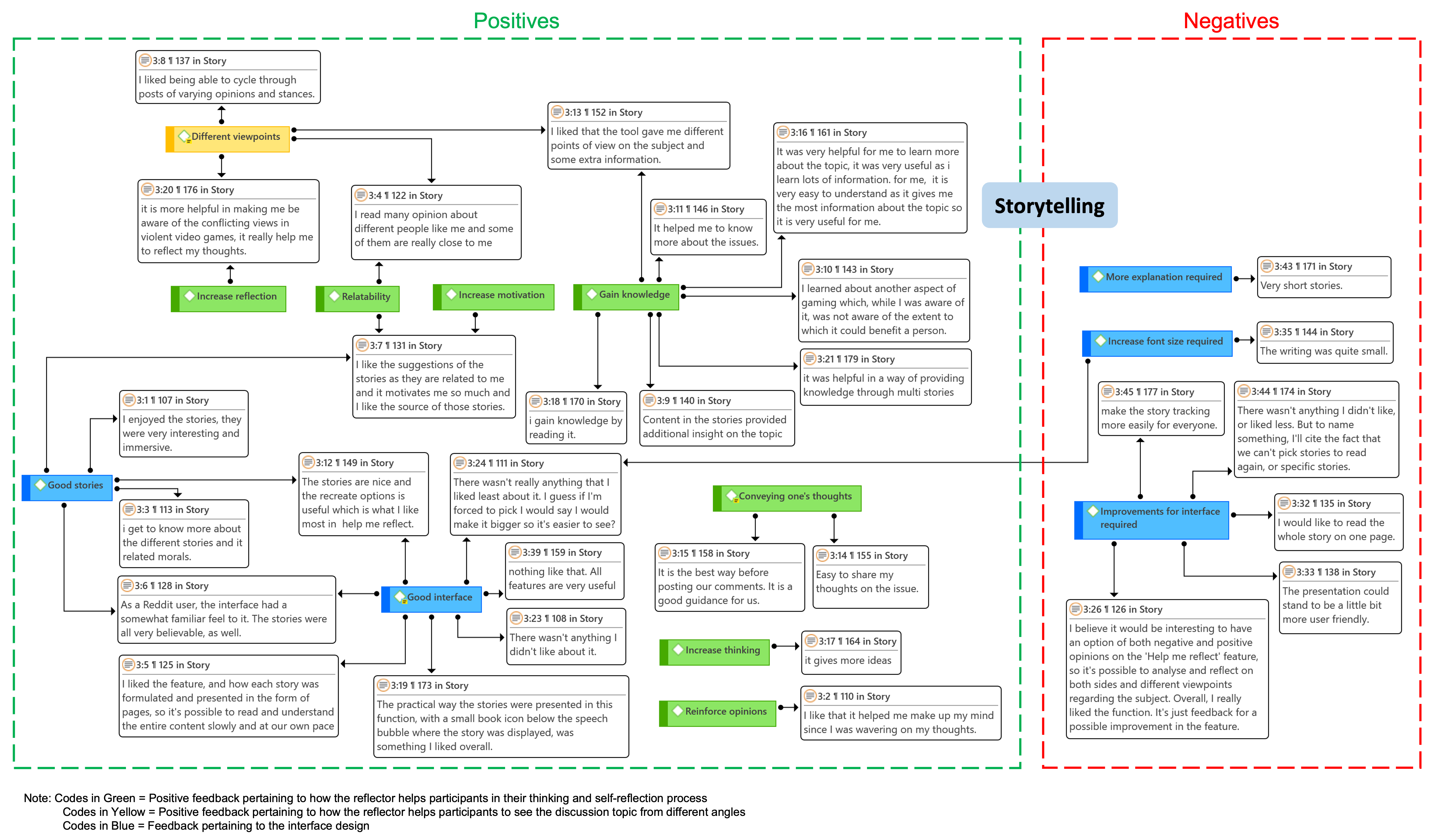}
  \caption{Participants' post-task feedback for the reflector - Storytelling. The figure provides a snapshot of the positive and negative aspects the reflector brought to the participants' experiences during their internal reflection process. Throughout the coding process, multiple codes could be applied to each participant's feedback.}
  \label{fig: storytelling post-task}
  \Description{Participants' Feedback on the storytelling reflector. We present an overview of participants' feedback regarding the storytelling reflector. Feedback are categorized into two main sections: Positives and Negatives, each represented with distinctive color codes for clarity. Green: Positive feedback highlighting how the storytelling Reflector assists participants in their thinking and self-reflection process. Yellow: Positive feedback indicating how the storytelling reflector helps participants explore the discussion topic from various perspectives. Blue: Feedback related to the interface design of the storytelling Reflector. Red: Negative feedback pointing out issues or challenges participants encountered while using the storytelling Reflector.}
\end{figure*}

\begin{table*}[!htbp]
\caption{Full List of the Post-Task Feedback for Storytelling: The numbers represent the frequency of occurrence of specific feedback mentioned by participants. The feedback are listed in descending order, from the highest to the lowest.}
\label{tab: storytelling post-task feedback}
\begin{tabular}{@{}cccc@{}}
\toprule
\multicolumn{2}{c}{\textbf{Positives}} & \multicolumn{2}{c}{\textbf{Negatives}} \\ 
\midrule
Good interface & 17 & Improvements for interface & 8 \\
Knowledge acquisition & 7 & More explanation & 3 \\
Good stories & 6 & Increase font size & 1 \\
Different viewpoints & 4 &  &  \\
Relatability & 2 &  &  \\
Increase reflection & 2 &  &  \\
Conveying one's thoughts & 2 &  &  \\
Increase motivation & 1 &  &  \\
Heightened thinking & 1 &  &  \\
Reinforce opinions & 1 &  &  \\
\bottomrule
\end{tabular}
\Description{This table is an extension of the top three post-task feedback for the storytelling reflector. We present the full List of the post-task feedback, both positives and negatives feedback for storytelling. We also report the numbers representing the frequency of occurrence of specific feedback mentioned by participants. The feedback are listed in descending order, from the highest to the lowest occurrence.}
\end{table*}

\newpage
\begin{figure*}[!htbp]
  \centering
  \includegraphics[scale=0.42]{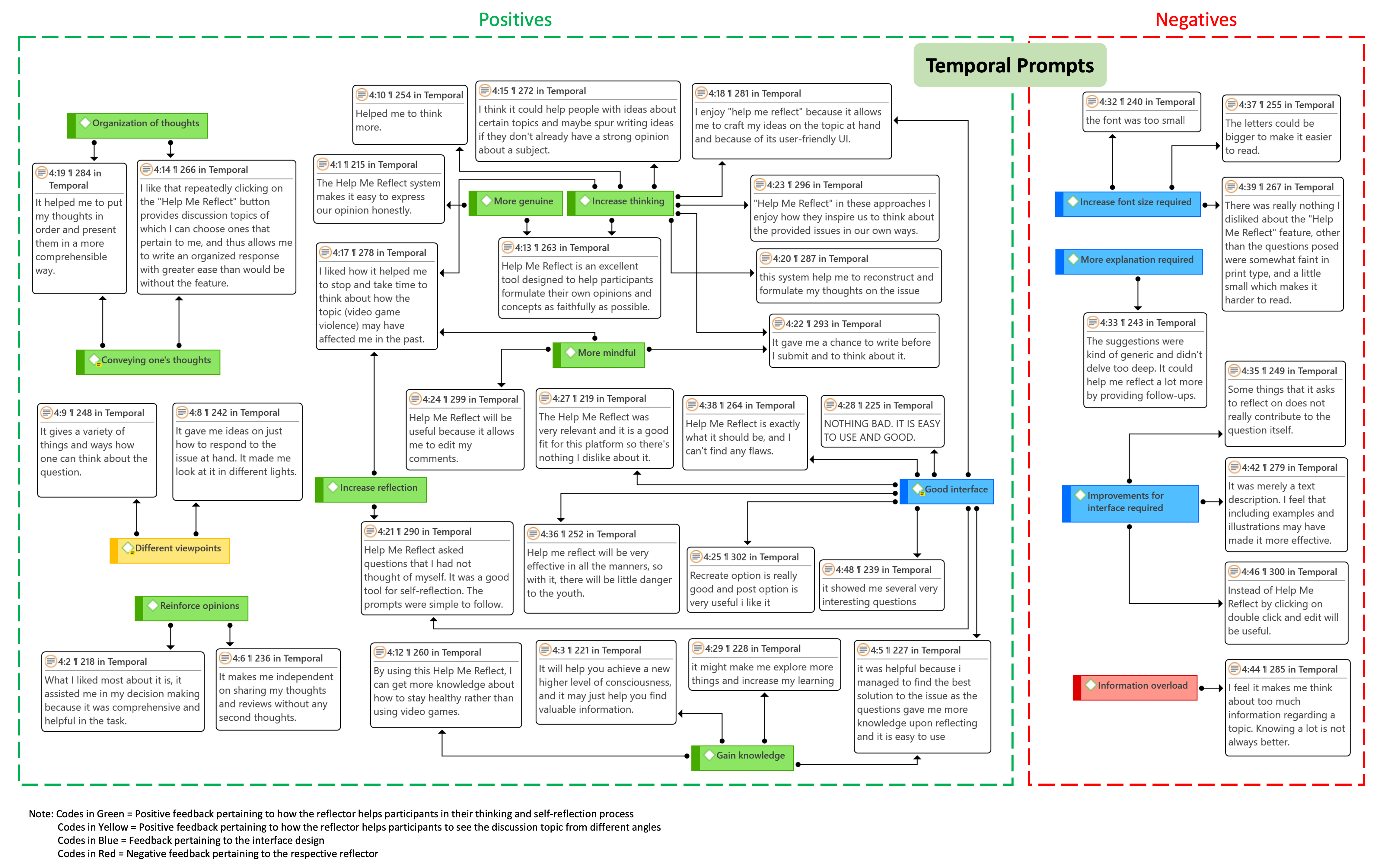}
  \caption{Participants' post-task feedback for the reflector - Temporal Prompts. The figure provides a snapshot of the positive and negative aspects the reflector brought to the participants' experiences during their internal reflection process. Throughout the coding process, multiple codes could be applied to each participant's feedback.}
  \label{fig: temporal prompts post-task}
  \Description{Participants' Feedback on the temporal prompts reflector. We present an overview of participants' feedback regarding the temporal prompts Reflector. Feedback are categorized into two main sections: Positives and Negatives, each represented with distinctive color codes for clarity. Green: Positive feedback highlighting how the temporal prompts reflector assists participants in their thinking and self-reflection process. Yellow: Positive feedback indicating how the temporal prompts reflector helps participants explore the discussion topic from various perspectives. Blue: Feedback related to the interface design of the temporal prompts reflector. Red: Negative feedback pointing out issues or challenges participants encountered while using the temporal prompts reflector.}
\end{figure*}

\begin{table*}[!htbp]
\caption{Full List of the Post-Task Feedback for Temporal Prompts: The numbers represent the frequency of occurrence of specific feedback mentioned by participants. The feedback are listed in descending order, from the highest to the lowest.}
\label{tab: temporal prompts post-task feedback}
\begin{tabular}{@{}cccc@{}}
\toprule
\multicolumn{2}{c}{\textbf{Positives}} & \multicolumn{2}{c}{\textbf{Negatives}} \\ 
\midrule
Good interface & 18 & Increase font size & 3 \\
Heightened thinking & 9 & Improvements for interface & 3 \\
Knowledge acquisition & 5 & More explanation & 1 \\
More mindful & 3 & Information overload & 1 \\
More genuine & 2 &  &  \\
Reinforce opinions & 2 &  & \\
Organization of thoughts & 2 &  &  \\
Conveying one's thoughts & 2 &  &  \\
Different viewpoints & 2 &  &  \\
Increase reflection & 1 &  &  \\
\bottomrule
\end{tabular}
\Description{This table is an extension of the top three post-task feedback for the temporal prompts reflector. We present the full List of the post-task feedback, both positives and negatives feedback for temporal prompts. We also report the numbers representing the frequency of occurrence of specific feedback mentioned by participants. The feedback are listed in descending order, from the highest to the lowest occurrence.}
\end{table*}

\newpage
\label{appendix: summary results study 2}


\begin{figure*}[!htbp]
  \centering
  \includegraphics[width=0.8\linewidth]{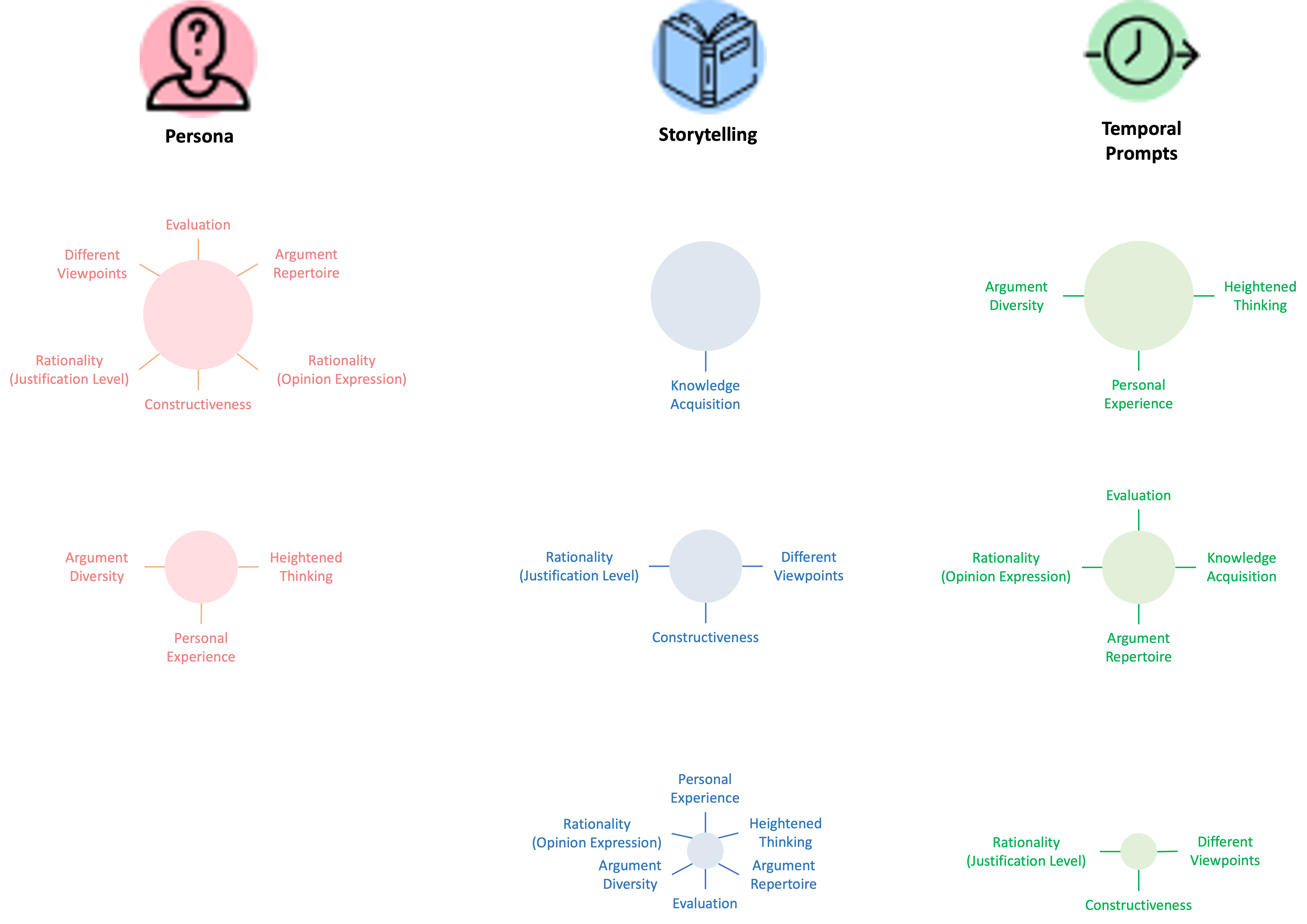}
  \caption{Visual Summary of the Results in Study 2. The results (quantitative results, qualitative results from content analysis and qualitative results from post-task feedback) for study 2 are summarized here. In this visual representation, we utilize circle size to convey the relative performance of each reflector across various deliberative dimensions when compared to the other two reflectors, with larger circles indicating better performance.}
  \label{fig: summary of results study 2}
  \Description{In our visual summary for Study 2, we employ a graphical representation featuring three circles for each of the reflectors: persona, storytelling, and temporal prompts. The circles vary in size, with the largest at the top, followed by a medium-sized circle, and the smallest at the bottom in a vertical arrangement. This visual representation utilizes the circle size to convey the relative performance of each reflector across various deliberative dimensions when compared to the other two reflectors. For instance, the largest circle representing persona encompasses the deliberative dimensions of evaluation, argument repertoire, rationality (opinion expression), constructiveness, rationality (justification level), and different viewpoints. This indicates that persona excels in these dimensions in comparison to storytelling and temporal prompts.}
\end{figure*}

\newpage
\label{appendix: reflective category table}

\begin{table*}[!htbp]
\caption{Table of Reflective Prompts. According to King~\cite{king1994inquiry}, inquiry-based and open-ended prompts using the following nature of question stems have the ability to facilitate higher order critical thinking and reflection as it requires people to reflect upon and reconcile various perspectives and solutions.}
\label{tab: reflective category table}
\centering
\scalebox{0.69}{
\begin{tblr}{
  width = \linewidth,
  colspec = {Q[300]Q[300]Q[600]},
  cell{2}{1} = {r=3}{},
  cell{2}{3} = {r=3}{},
  cell{5}{1} = {r=3}{},
  cell{5}{3} = {r=4}{},
  cell{11}{1} = {r=2}{},
  cell{11}{3} = {r=2}{},
  vlines,
  hline{1-2,5,9-11,13-16} = {-}{},
  hline{3-4,6-7,12} = {2}{},
  hline{8} = {1-2}{},
}
\textbf{Reflective Category/ Reflective Thinking Skill} & \textbf{Nature of Question Stems/ Textual Prompts} & \textbf{What it does}\\
Application & What is a new example of…? & Implementation or use of violent video games. It prompts individuals to think about scenarios where violent video games can be applied and encourages them to consider practical applications. This type of reflective prompts fosters a deeper understanding to the subject matter.\\ & How could … be used to…? & \\ & How does … apply to everyday life? & \\ 
Analysis/Inference & What are the implications of…? & Examining and drawing conclusions or inferences from the prompts. It encourages individuals to think about the broader consequences or potential outcomes, both positive and negative arising due to video games. This type of reflective prompts encourages a deeper exploration of the underlying reasons and mechanisms, fostering critical thinking and more comprehensive understanding of the subject matter. \\ & Explain why … & \\ & Explain how … & \\ Analysis of significance & Why is … important? & \\
Identification and Creation of Analogies and Metaphors & What is … analogous to…? & Draw comparisons between the topic to familiar concepts in order to illustrate similarities or to convey a deeper understanding. This type of reflective prompt encourages creative thinking and helps convey complex ideas in a more accessible and relatable manner.\\
Activation of prior knowledge & What do we already know about…? & Stimulates the information and understanding of what an individual already possesses. It prompts them to recall and use their existing knowledge or experiences to relate to the topic. This type of reflective prompts connect new information to what is already known, facilitating a deeper understanding and more effective processing of the topic.\\
Analysis of relationship (cause-effect) & How does … affect…? & Prompts individuals to consider the cause-and-effect relationship between two elements or variables. It encourages the exploration of the consequences that result from video games. This type of reflective prompts stimulates thinking about the interconnectedness and influence of different elements in a scenario. It is useful in understanding the relationships and dynamics between various factors and gaining insights into the broader implications of the subject matter.\\ & What do you think causes…? Why? & \\
Synthesis of ideas & What are some possible solutions to the problem of…? & Prompts individuals to engage in creative thinking and combine different ideas to generate potential solutions for a given problem. It encourages the exploration of innovative approaches, strategies or interventions that could address the identified issue. This type of reflective prompts foster a problem-solving mindset and draw on the knowledge of the individual to propose practical and effective solutions.\\
Evaluation and provision of evidence & Do you agree or disagree with this statement: …? & Prompts individuals to assess a claim. It encourages them to articulate their rationale while expressing their agreements or disagreements. This type of reflective prompts promote critical thinking, analysis and the ability to express well-supported opinions, fostering a deeper exploration of the subject matter.\\
Taking other perspectives & How do you think … would see the issue of…? & Prompts individuals to consider how someone else might perceive the subject matter. This type of reflective prompts aims to engage individuals in perspective-taking through imagining the viewpoints and considerations of others. It fosters empathy and broaden one's understanding on the subject matter.
\end{tblr}}
\Description{This table details the question stems and the characteristics of prompts following King's framework. It comprises three columns with headers: Reflective Category/Reflective Thinking Skills, Nature of Question Stems/Textual Prompts, and What It Does. The Reflective Category column categorizes the reflective thinking skill targeted by the question stems. The Nature of Question Stems column lists various question stems and textual prompts associated with the respective reflective thinking skill. The What It Does column details the content of the Nature of Question Stems column, explaining how it aims to stimulate the specific reflective thinking skill.}
\end{table*}

\end{document}